%% file: structuralcorrelations.tex
\documentclass[a4paper,10pt]{article}

\usepackage{amsfonts}
\usepackage{amsmath}

\usepackage{tikz}
\usepackage{graphicx}
\usepackage{epstopdf}
\usepackage{subcaption}

\usepackage{multirow}

\usepackage{a4wide}

\newcommand{\la}{\left\langle}
\newcommand{\ra}{\right\rangle}

\begin{document}

\title{Phase transitions for scaling of structural correlations in directed networks.}
\author{Pim van der Hoorn\footnote{University of Twente, w.l.f.vanderhoorn@utwente.nl}, 
	Nelly Litvak\footnote{University of Twente, n.litvak@utwente.nl}}
\date{\today}

\maketitle

\begin{abstract}
	Analysis of degree-degree dependencies in complex networks, and their impact on processes on 
	networks requires null models, i.e. models that generate uncorrelated scale-free networks. Most 
	models to date however show structural negative dependencies, caused by finite size effects. We 
	analyze the behavior of these structural negative degree-degree dependencies, using rank based 
	correlation measures, in the directed Erased Configuration Model. We obtain expressions for the 
	scaling as a function of the exponents of the distributions. Moreover, we show that this scaling 
	undergoes a phase transition, where one region exhibits scaling related to the natural cut-off 
	of the network while another region has scaling similar to the structural cut-off for 
	uncorrelated networks. By establishing the speed of convergence of these structural dependencies 
	we are able to asses statistical significance of degree-degree dependencies on finite complex 
	networks when compared to networks generated by the directed Erased Configuration Model. 
\end{abstract}

\input{introduction}

\input{directednetworks}

\input{directed_ecm}

\input{degree_dependencies_ecm}

\input{analysis_correlations}

\input{scaling_other_cases}

\input{conclusion}

\noindent\textbf{Acknowledgments:} \\
	All computations in this paper where done using the \emph{fastutil} package and the WebGraph framework,
	\cite{Boldi2004}, from the Laboratory for Web Algorithmics http://law.di.unimi.it/software.php.
	This work is supported by the EU-FET Open grant NADINE (288956).

\bibliographystyle{plain}
\bibliography{bib/structuralcorrelations}

\end{document}

%% file: introduction.tex
\section{Introduction}

The tendency of nodes in a network to be connected to nodes of similar large or small degree, called 
network assortativity, degree mixing or degree-degree dependency, is an important characterization of 
the topology of the network, influencing many processes on the network. It has received significant 
attention in the literature, for instance in the field of network stability~\cite{Vazquez2003}, 
attacks on P2P networks~\cite{srivastava2011} and epidemics~\cite{Boguna2002,Boguna2003a}.

An important method to analyze these degree-degree dependencies or their influence on other network 
properties or processes on the network, is to compare results to an average over several instances of 
similar networks with neutral mixing. These null models often come in two flavors. The first approach 
is to sample from graphs with the same degree sequence but neutral mixing. A widely accepted 
methodology for such sampling is through the local rewiring model, \cite{Maslov2002}, which takes the 
original network and randomly swaps edges until a randomized version is attained. The disadvantage of 
these methods is that they have no theoretical performance guarantees. The second approach is to 
generate a random graph with neutral mixing, which preserves basic features, such as the degree 
distribution. A well known model of this type is the Configuration Model (CM)~\cite{Bollobas1980, 
Molloy1995, Newman2001}. Here the degrees of vertices are drawn independently from the given 
distribution, under the restriction that the total sum of degrees is even. Then the stubs are paired 
uniformly at random to form edges. If we want to obtain a simple graph in this way, we can either 
rewire till a simple graph is generated (Repeated Configuration Model), or we remove the excess 
edges and self loops (Erased Configuration Model).

We note that there are many other methods, that generate simple random graphs and have 
theoretically established performance guarantees. For example, sequential algorithms based on the 
properties of graphical sequences were proposed for undirected networks
\cite{Blitzstein2011sequential,DelGenio2010} and directed networks~\cite{Kim2012}. Another example 
is a grand-canonical model in~\cite{Squartini2011} that generates a graph with given average 
degrees using a maximum-entropy method. However, to the best of our knowledge, none of these 
methods has an efficient implementation. Even the complexity $O(N E)$ in
\cite{DelGenio2010,Kim2012}, where $N$ is the network size and $E$ the number of edges, is arguably 
not feasible for truly large networks, such as Wikipedia or Twitter.

Although for both local rewiring and the Configuration Model neutral mixing is expected, since there 
is no preference in connecting two vertices, negative correlations are observed,~\cite{Catanzaro2005, 
Maslov2004,Park2003}, for scale-free networks with infinite variance of degrees, i.e. where the degree 
distribution satisfies 
\begin{equation}
	P(k) \sim k^{-(\gamma + 1)}, \quad 1 < \gamma \le 2.
\label{eq:scalefree_distribution}
\end{equation} 
In~\cite{Maslov2004} this phenomenon is explained by observing that if one allows at most one edge 
between two vertices, nodes with large degree must connect to nodes of small degree because there are 
simply not enough distinct large nodes to connect to. A similar explanation is given in
\cite{Catanzaro2005}. Here, however, this is then related to the difference in scaling between the 
\emph{natural} and \emph{structural cut-off} of the network. The former is defined 
\cite{Dorogovtsev2002} as the degree value $k_c$, of which, on average, only one instance is observed:
\begin{equation}
	N \int_{k_c}^\infty P(k) dk \sim 1.
	\label{eq:natural_cutoff}
\end{equation}
The structural cut-off is defined as the value $k_s$ for which the ratio between the average number
of edges that connect any two vertices of degree $k_s$, and the maximum possible number of such edges 
in a simple graph, is $1$. For networks with degree distribution~(\ref{eq:scalefree_distribution}) it 
follows from~(\ref{eq:natural_cutoff}) that the natural cut-off scales as $N^{1/\gamma}$, while the 
structural cut-off for uncorrelated networks scales as, see~\cite{Boguna2004}, $N^{1/2}$. Therefore, 
when $\gamma < 2$, the natural cut-off scales at a slower rate which in turn gives rise to structural 
negative correlations.

To remedy these finite size effects the authors of~\cite{Catanzaro2005} propose an Uncorrelated 
Configuration Model. This model follows the same procedure as the regular Configuration Model, with 
the addition that the sampled degrees are bounded, $m \le k_i \le N^{1/2}$. Experiments in
\cite{Catanzaro2005} indeed show that these networks are uncorrelated. However, many scale-free 
networks, for instance Twitter, have nodes who's degree is of larger order than $N^{1/2}$, which is 
a characteristic property of scale-free graphs. For example, Table~\ref{tbl:wiki_degree_stats} 
displays the characteristics of Wikipedia networks for different languages. Here we see that the 
maximum out-degree could be considered to be of order $N^{1/2}$, while the maximum in-degree is 
definitely of a much larger scale. Therefore, randomized versions of these networks, generated by 
the Uncorrelated Configuration Model, do not have the same basic degree characteristics as the 
original network, since the maximum degree is restricted. Hence, they are less suitable for 
comparison of the degree-degree dependencies.

\input{tables_wiki_degree_stats}

In this paper we consider the directed Erased Configuration Model (ECM), \cite{chen2013}, where 
after the pairing self loops are removed and multiple edges are merged. In our recent work
\cite{Hoorn2014}, Section 5, we showed that this model has neutral mixing in the infinite network 
size limit. The idea behind this result is that the total average number of erased edges per node, 
which defines the difference in the correlations between the CM and the ECM, goes to zero when the 
size of the network grows. By this result, from a purely mathematical point of view the ECM is a 
null model for degree-degree dependencies in the limit. Moreover, asymptotically, the degree 
distributions are preserved and hence, all basic degree characteristics. Still, for finite sizes, 
structural dependencies are present. 

Rather than trying to control these correlations, our goal is to evaluate their magnitude and 
investigate their size dependence. We obtain the scaling for the structural correlations in the ECM, 
in terms of the power law exponents of the in- and out-degrees. In particular, we show that this 
scaling undergoes an interesting phase transition, and can be dominated by terms related to either 
the structural or the natural cut-off of the network. To the best of our knowledge, this is the 
first study that provides a systematic mathematical characterization for the magnitude of negative 
correlations in a simple graph with neutral mixing.

By determining the scaling of the structural correlations we can asses the significance of measured 
correlations as well as their influence on network processes, on real world networks of finite size, 
by comparing them to the directed Erased Configuration Model. This approach has the advantage of 
preserving the degree characteristics of the original network, it can be easily implemented and 
applied to all networks with scale-free degree distributions and finite expectation.

%% file: tables_wiki_degree_stats.tex
\begin{table}%
\centering
\begin{tabular}{c|c|c|c|c|c|c}
	Wikipedia & $N$ & $N^{1/2}$& $\gamma_+$ & $\gamma_-$ & $\max D^+$ & $\max D^-$ \\
	\hline
	DE & 1,532,978 & 1,238 &1.80 & 1.05 & 5,032 & 118,064 \\
	EN & 4,212,493 &2,052 &2.14 & 1.20 & 8,104 & 432,629 \\
	IT & 1,017,953 & 1,009& 1.96 & 1.05 & 5,212 & 91,588 \\
	NL & 1,144,615 & 1,070&1.82 & 1.10 & 10,175 & 102,450 \\
	PL & 949,153 &974 &1.90 & 1.04 & 4,100 & 112,537
\end{tabular}
\caption{Basic degree characteristics of Wikipedia networks. The exponents of 
the degree distributions are estimated using the implementation of the techniques 
from~\cite{Clauset2009} by Peter Bloem, http://github.com/Data2Semantics/powerlaws.}
\label{tbl:wiki_degree_stats}
\end{table}

%% file: directednetworks.tex
\begin{figure}[t]
	\centering
	\input{figures_fourdependencies}
	\caption{The four different degree-degree dependency types in directed networks.}%
	\label{fig:four_dependency_types}%
\end{figure}
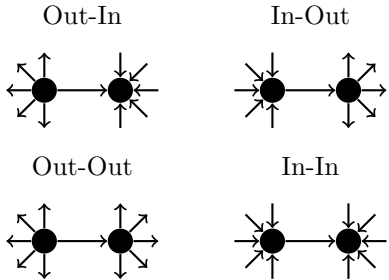

\begin{figure}[t]%
	\begin{subfigure}{.32\linewidth}
		\includegraphics[scale=0.35]{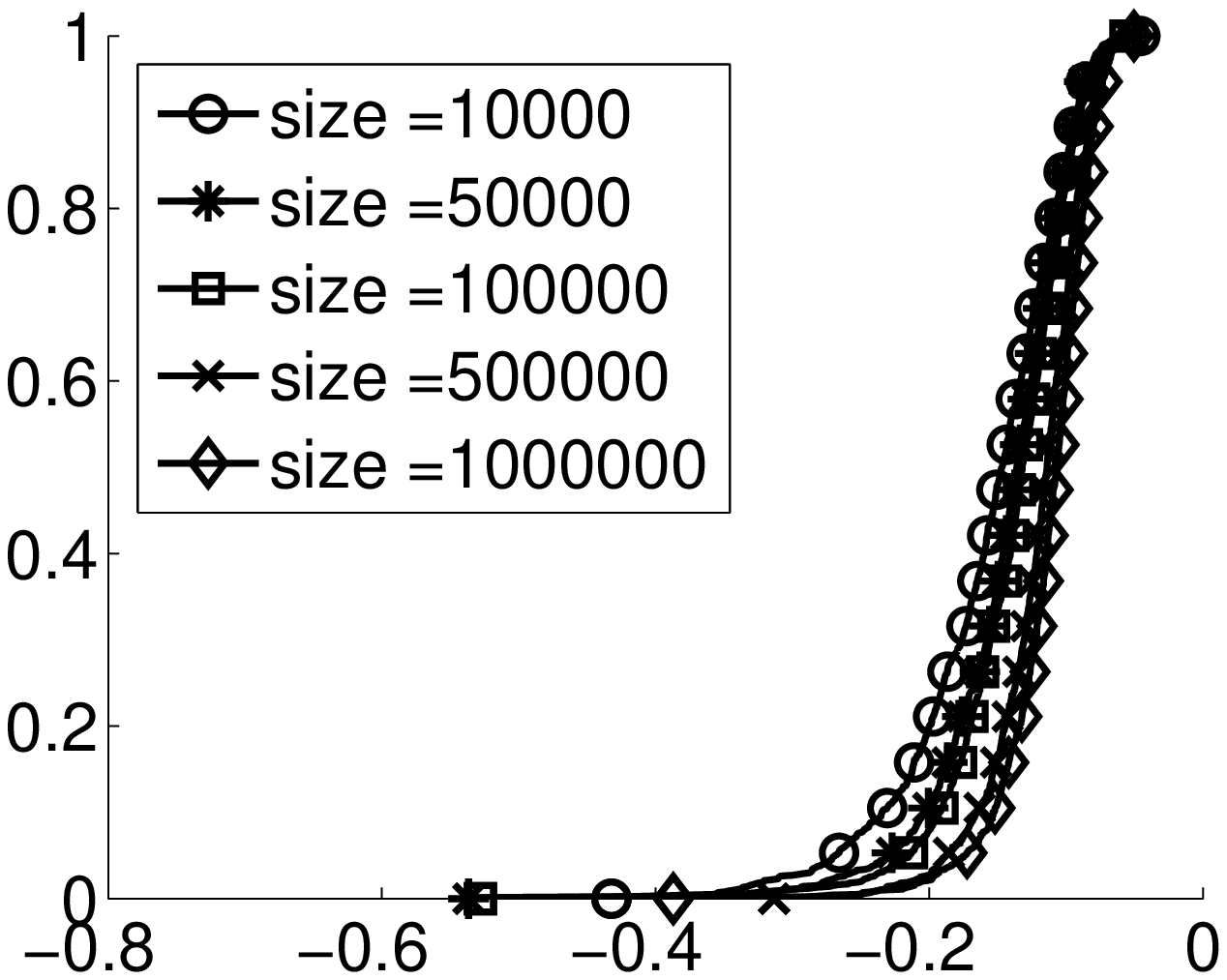}
		\caption{$\rho_+^-$}
		\label{sfig:spearman_uniform_outin}
	\end{subfigure}
	\begin{subfigure}{.32\linewidth}
		\includegraphics[scale=0.35]{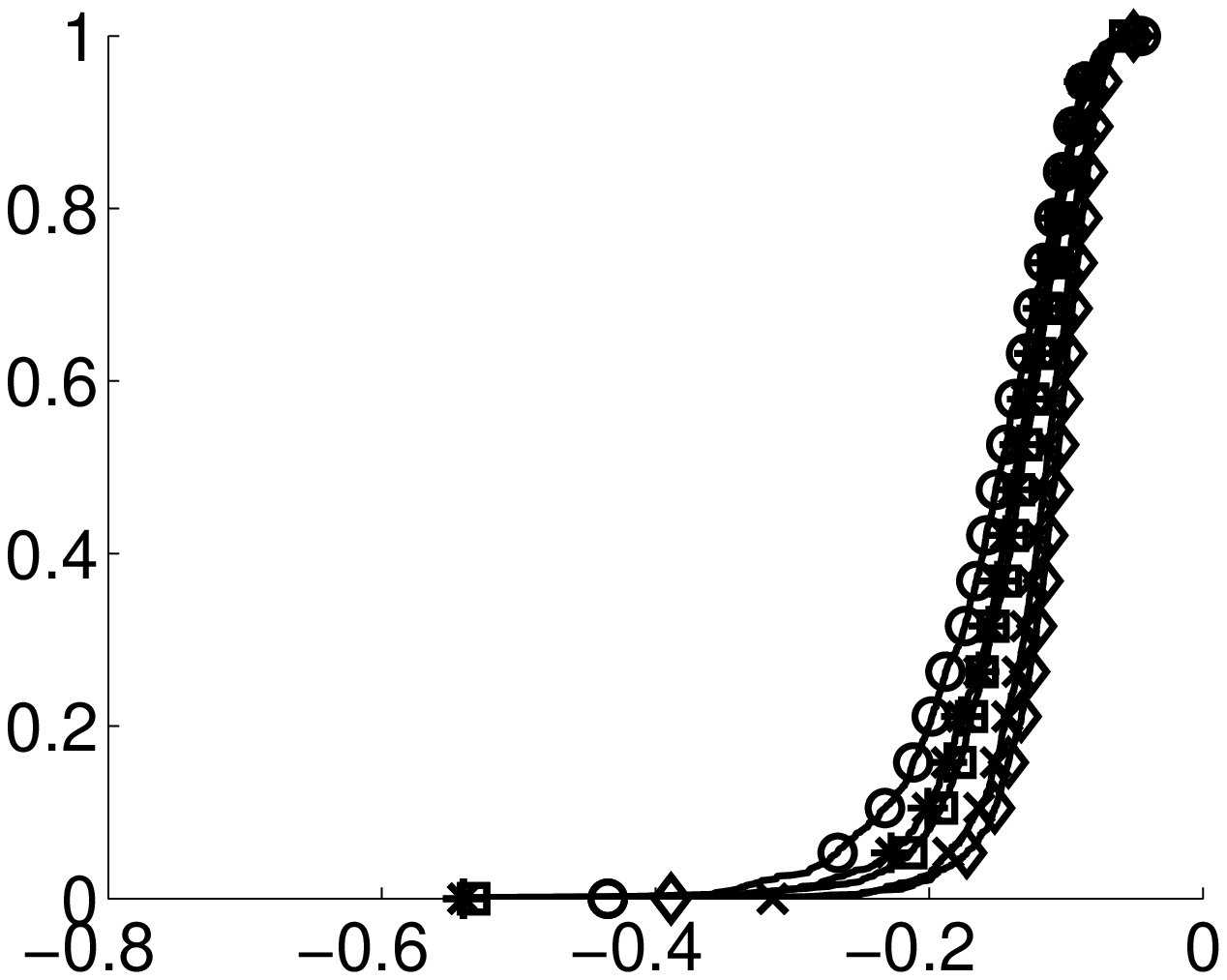}
		\caption{$\overline{\rho}_+^-$}
		\label{sfig:spearman_average_outin}
	\end{subfigure}
	\begin{subfigure}{.32\linewidth}
		\includegraphics[scale=0.35]{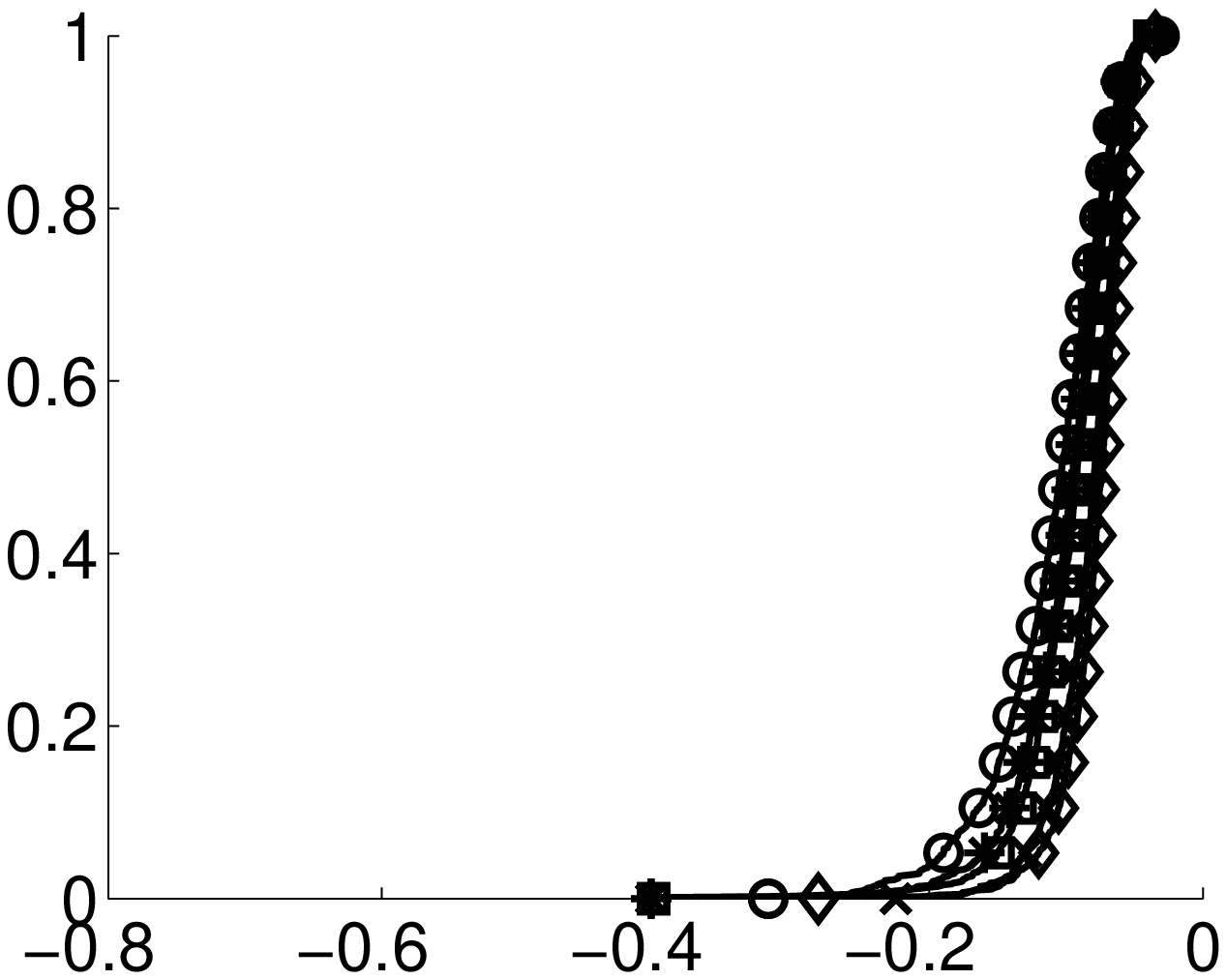}
		\caption{$\tau_+^-$}
		\label{sfig:kendall_outin}
	\end{subfigure}
	\caption{Plots of the empirical cumulative distribution of $\rho_+^-$, $\overline{\rho}_+^-$ and
	$\tau_+^-$ for ECM graphs of different sizes with $\gamma_\pm = 1.2$. Each plot is based
	on $10^3$ realizations of the model.}
	\label{fig:similar_measures_example}%
\end{figure}

\begin{figure}
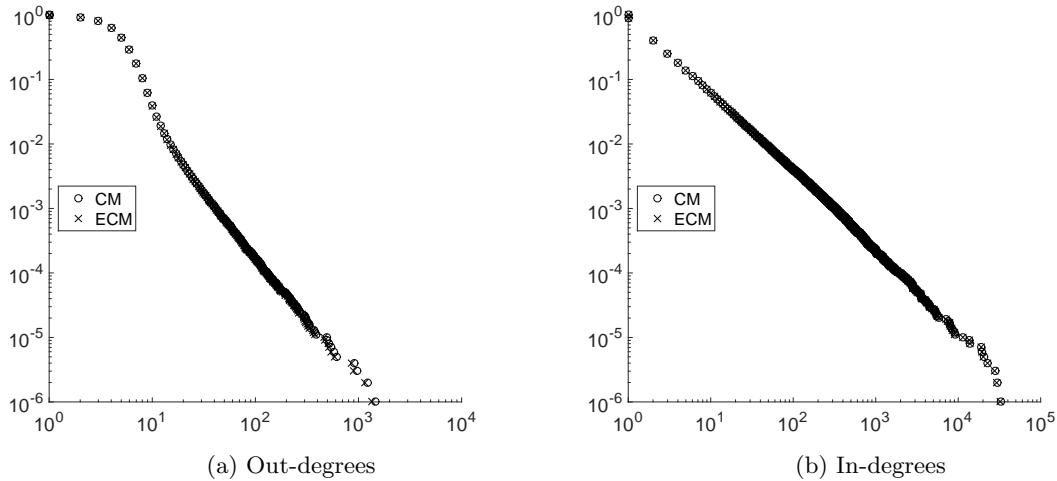

	\begin{subfigure}{0.5\linewidth}
		\includegraphics[scale=0.35]{figures_degrees_1219_outdegrees.eps}
		\caption{Out-degrees}
	\end{subfigure}~
	\begin{subfigure}{0.5\linewidth}
		\includegraphics[scale=0.35]{figures_degrees_1219_indegrees.eps}
		\caption{In-degrees}
	\end{subfigure}
	\caption{Plots of the out- and in-degree distribution, on log-log scale, for a graph generated by 
		the ECM, of size $10^6$ with $\gamma_+ = 1.9$ and $\gamma_- = 1.2$, before (CM) and after (ECM) 
		the removing of edges.}
	\label{fig:distributions_cm_ecm}
\end{figure}

\section{Degree-degree dependencies in random directed networks}

We analyze degree-degree dependencies in random directed networks of size $N$, where the distribution 
of the out- and in-degree ($D^+$, $D^-$) follow, respectively,
\begin{equation}
	P^+(k) \sim k^{-(\gamma_+ + 1)} \hspace{3pt} \text{and} \hspace{3pt} P^-(\ell) \sim \ell^{-( 
	\gamma_- + 1)},	\hspace{3pt} \gamma_\pm > 1.
\label{eq:in_out_distribution}
\end{equation}

In directed networks one can consider four types of degree-degree dependencies, depending on the 
choice of the degree type on both sides of an edge, see Figure~\ref{fig:four_dependency_types}. For 
the remainder of this paper we denote by $E$ the number of edges and adopt the notation style from
\cite{Foster2010,Hoorn2014a} to index the degree types by $\alpha, \beta \in \{+, -\}$.

A common measure for degree-degree dependencies, introduced in~\cite{Newman2002}, computes Pearson's 
correlation coefficients on the joint data $(D_i^\alpha, D_j^\beta)_{i \to j}$, where the indices 
run over all $i, j$ for which there is an edge $i \to j$. 

However, Pearson's correlation coefficients are unable to measure strong negative degree-degree 
dependencies in large networks where the variance of the degrees is infinite, as was shown for 
undirected networks in~\cite{Litvak2013, Dorogovtsev2010} and for directed networks in
\cite{Hoorn2014a}. Since our interest is mainly in networks in the infinite variance domain, i.e. 
$1 < \gamma_\pm \le 2$, we need different measures. In~\cite{Hoorn2014a} it was suggested to use 
rank correlations, related to Spearman's rho~\cite{Spearman1904} and Kendall's tau~\cite{Kendall1938}, 
to measure degree-degree dependencies. 

Spearman's rho computes Pearson's correlation coefficient on the ranks of $(D_i^\alpha, D_j^\beta)
_{i \to j}$ rather then their actual values. Since this data will contain many ties, one needs to use
ranking schemes that deal with these ties. In~\cite{Hoorn2014a} two such schemes are considered, 
resolving ties at random and assigning an average rank to tied values, which give two correlation
measures denoted by $\rho_\alpha^\beta$ and $\overline{\rho}_\alpha^\beta$, respectively. 
Here, the subscript index denotes the degree type of the source, while the superscript index denotes
the degree type of the target of a directed edge. For instance, $\rho_+^-$ denotes Spearman's rho for 
the Out-In dependency. The second rank correlation measure, Kendall's tau $\tau_\alpha^\beta$, 
calculates the normalized number of swaps needed to match the ranks of the joint data. 

Exact formulas for these three measures, in terms of the degrees, are given in~\cite{Hoorn2014a}. In
\cite{Hoorn2014} formulas are given in terms of the empirical distributions of $D^\alpha$ and $D^\beta$ 
and their joint distribution, evaluated at $(D^\alpha_i, D^\beta_j)$ for an edge $i \to j$ selected 
uniformly at random. From these it follows that if the network has neutral mixing, then $\rho_\alpha
^\beta$ and $\tau_\alpha^\beta$ are similar, while $\rho_\alpha^\beta$ and $\overline{\rho}_\alpha
^\beta$ differ by a term of $O(1)$, which does not influence the scaling. To illustrate this we 
plotted the empirical cdf's of $\rho_+^-$, $\overline{\rho}_+^-$ and $\tau_+^-$ for a collection of 
ECM graphs in Figure~\ref{fig:similar_measures_example}; where we clearly observe the similar behavior 
of the three measures. Therefore, for the analysis of degree-degree dependencies, we will only 
consider $\rho_\alpha^\beta$, which corresponds to Spearman's rho where ties are resolved uniformly 
at random.

%% file: figures_fourdependencies.tex
\tikzstyle{vertex}=[fill, circle, minimum size=4pt]
\tikzstyle{edge}=[style=thick, color=black]

\begin{tikzpicture}
	
	\draw node[vertex] (out_in_left) at (-2, 1) {};
	\draw node[vertex] (out_in_right) at (-1, 1) {};
	
	\draw node at (-1.5,2) {Out-In};
	\draw [->][edge] (out_in_left) -- (out_in_right);
	
	\draw [->][edge] (out_in_left) -- (-2,1.5);
	\draw [->][edge] (out_in_left) -- (-2.354,1.354);
	\draw [->][edge] (out_in_left) -- (-2.5,1);
	\draw [->][edge] (out_in_left) -- (-2.354,0.646);
	\draw [->][edge] (out_in_left) -- (-2,0.5);
	
	\draw [->][edge] (-1,1.5) -- (out_in_right);
	\draw [->][edge] (-0.646,1.354) -- (out_in_right);
	\draw [->][edge] (-0.5,1) -- (out_in_right);
	\draw [->][edge] (-0.646,0.646) -- (out_in_right);
	\draw [->][edge] (-1,0.5) -- (out_in_right);
	
	\draw node[vertex] (in_out_left) at (1, 1) {};
	\draw node[vertex] (in_out_right) at (2, 1) {};
	
	\draw node at (1.5,2) {In-Out};
	\draw [->][edge] (in_out_left) -- (in_out_right);
	
	\draw [->][edge] (1,1.5) -- (in_out_left);
	\draw [->][edge] (0.646,1.354) -- (in_out_left);
	\draw [->][edge] (0.5,1) -- (in_out_left);
	\draw [->][edge] (0.646,0.646) -- (in_out_left);
	\draw [->][edge] (1,0.5) -- (in_out_left);
	
	\draw [->][edge] (in_out_right) -- (2,1.5);
	\draw [->][edge] (in_out_right) -- (2.354,1.354);
	\draw [->][edge] (in_out_right) -- (2.5,1);
	\draw [->][edge] (in_out_right) -- (2.354,0.646);
	\draw [->][edge] (in_out_right) -- (2,0.5);
	
	\draw node[vertex] (out_out_left) at (-2, -1) {};
	\draw node[vertex] (out_out_right) at (-1, -1) {};
	
	\draw node at (-1.5,0) {Out-Out};
	\draw [->][edge] (out_out_left) -- (out_out_right);
	
	\draw [->][edge] (out_out_left) -- (-2,-0.5);
	\draw [->][edge] (out_out_left) -- (-2.354,-0.646);
	\draw [->][edge] (out_out_left) -- (-2.5,-1);
	\draw [->][edge] (out_out_left) -- (-2.354,-1.354);
	\draw [->][edge] (out_out_left) -- (-2,-1.5);
	
	\draw [->][edge] (out_out_right) -- (-1,-0.5);
	\draw [->][edge] (out_out_right) -- (-0.646,-0.646);
	\draw [->][edge] (out_out_right) -- (-0.5,-1);
	\draw [->][edge] (out_out_right) -- (-0.646,-1.354);
	\draw [->][edge] (out_out_right) -- (-1,-1.5);
	
	\draw node[vertex] (in_in_left) at (1, -1) {};
	\draw node[vertex] (in_in_right) at (2, -1) {};
	
	\draw node at (1.5,0) {In-In};
	\draw [->][edge] (in_in_left) -- (in_in_right);
	
	\draw [->][edge] (1,-0.5) -- (in_in_left);
	\draw [->][edge] (0.646,-0.646) -- (in_in_left);
	\draw [->][edge] (0.5,-1) -- (in_in_left);
	\draw [->][edge] (0.646,-1.354) -- (in_in_left);
	\draw [->][edge] (1,-1.5) -- (in_in_left);
	
	\draw [->][edge] (2,-0.5) -- (in_in_right);
	\draw [->][edge] (2.354,-0.646) -- (in_in_right);
	\draw [->][edge] (2.5,-1) -- (in_in_right);
	\draw [->][edge] (2.354,-1.354) -- (in_in_right);
	\draw [->][edge] (2,-1.5) -- (in_in_right);
	
\end{tikzpicture}

%% file: directed_ecm.tex
\begin{figure}[t]%
	\begin{subfigure}{.5\linewidth}
		\includegraphics[scale=0.35]{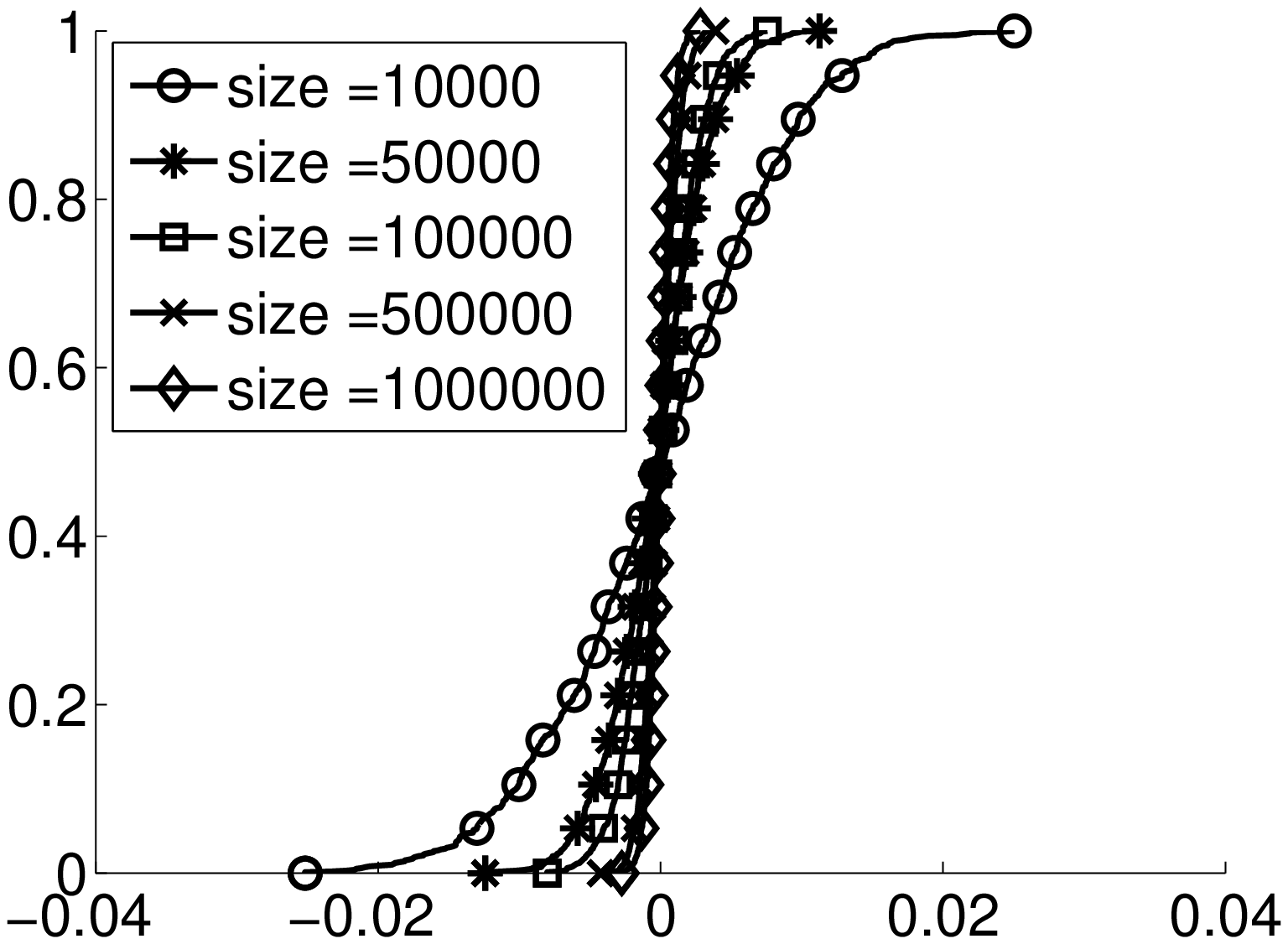}
		\caption{Out-In}
		\label{sfig:normal_corr_outin}
	\end{subfigure}~
	\begin{subfigure}{.5\linewidth}
		\includegraphics[scale=0.35]{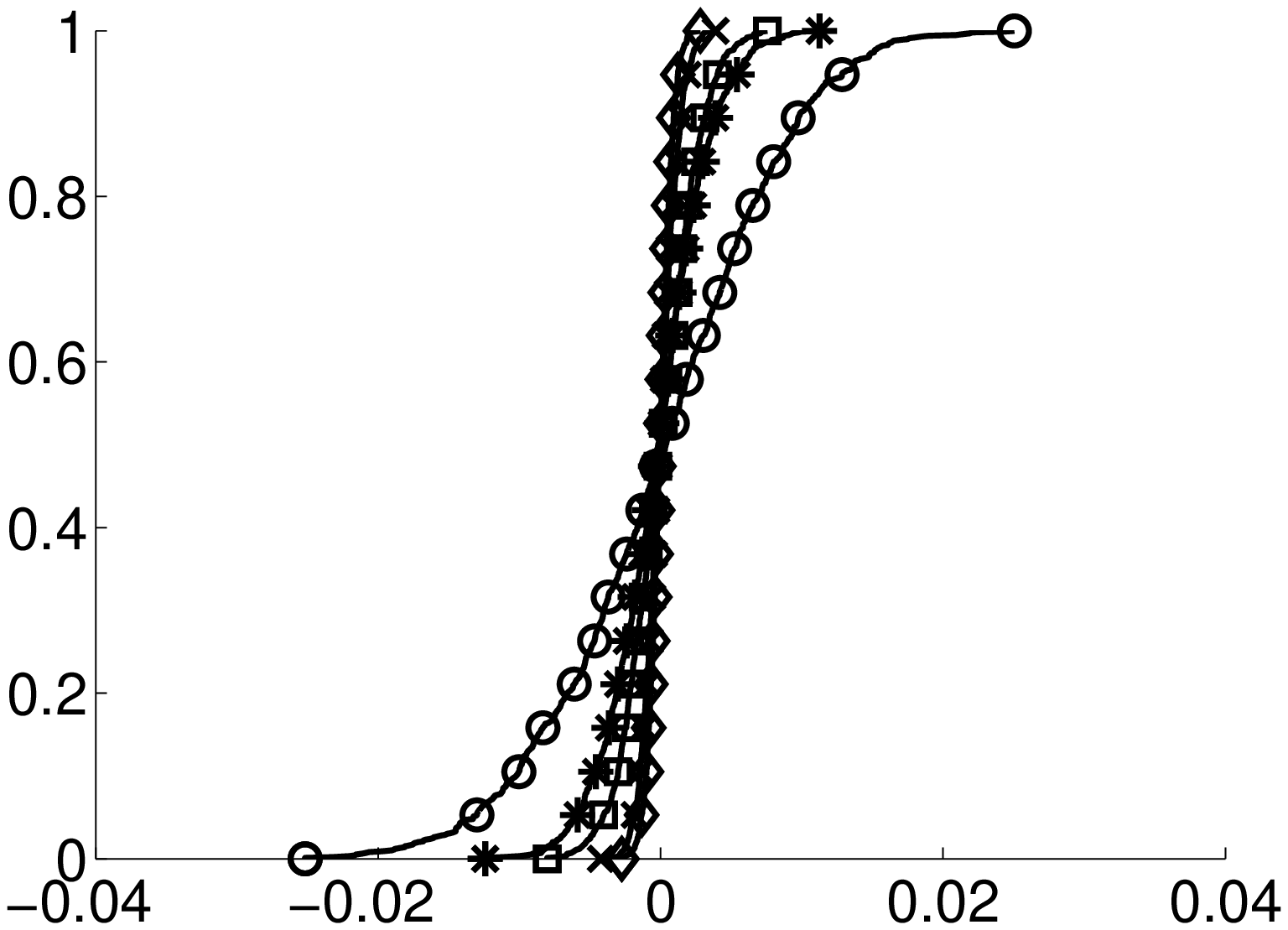}
		\caption{In-Out}
		\label{sfig:normal_corr_inout}
	\end{subfigure}
	
	\begin{subfigure}{.5\linewidth}
		\includegraphics[scale=0.35]{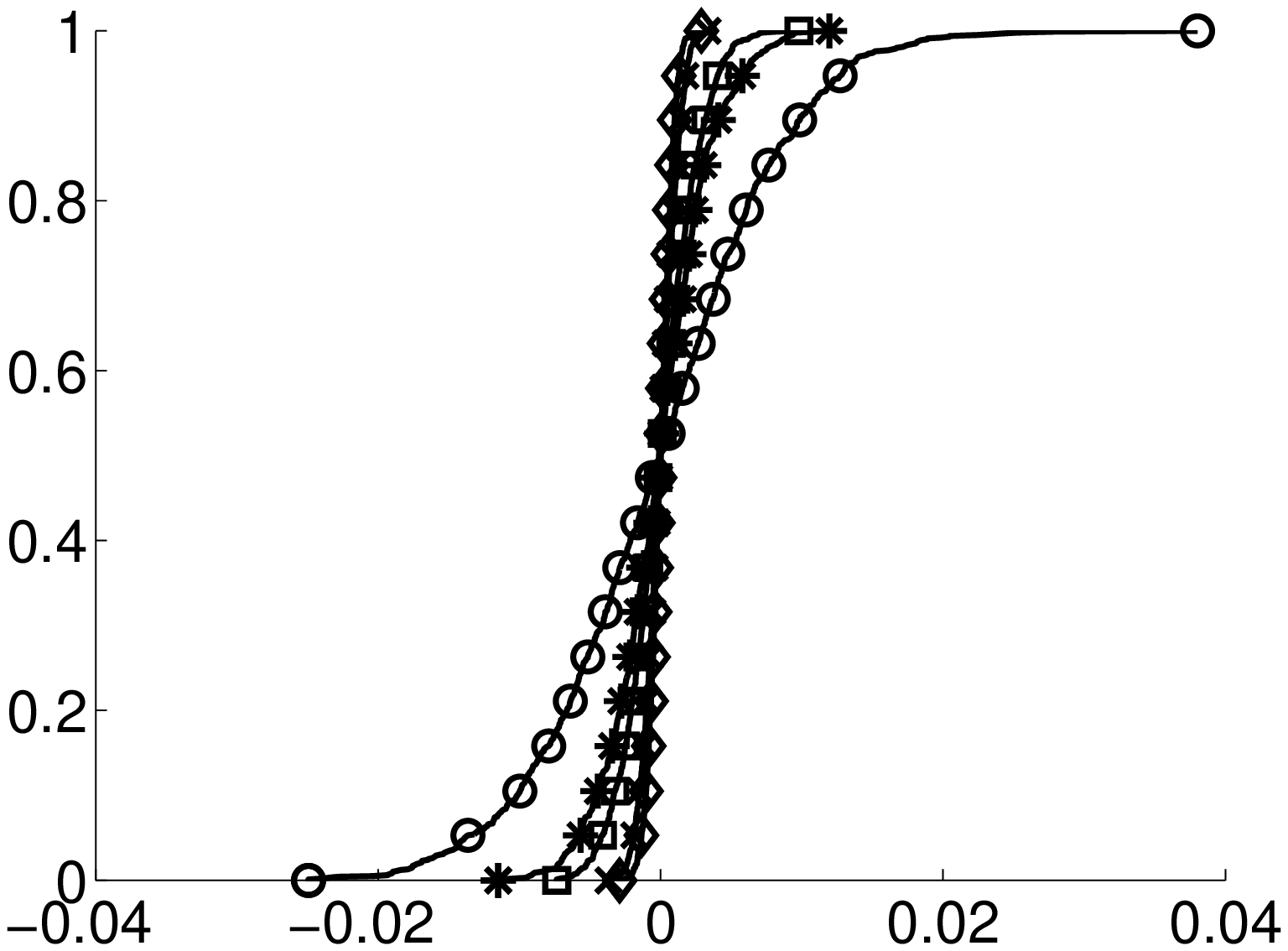}
		\caption{Out-Out}
		\label{sfig:normal_corr_outout}
	\end{subfigure}~
	\begin{subfigure}{.5\linewidth}
		\includegraphics[scale=0.35]{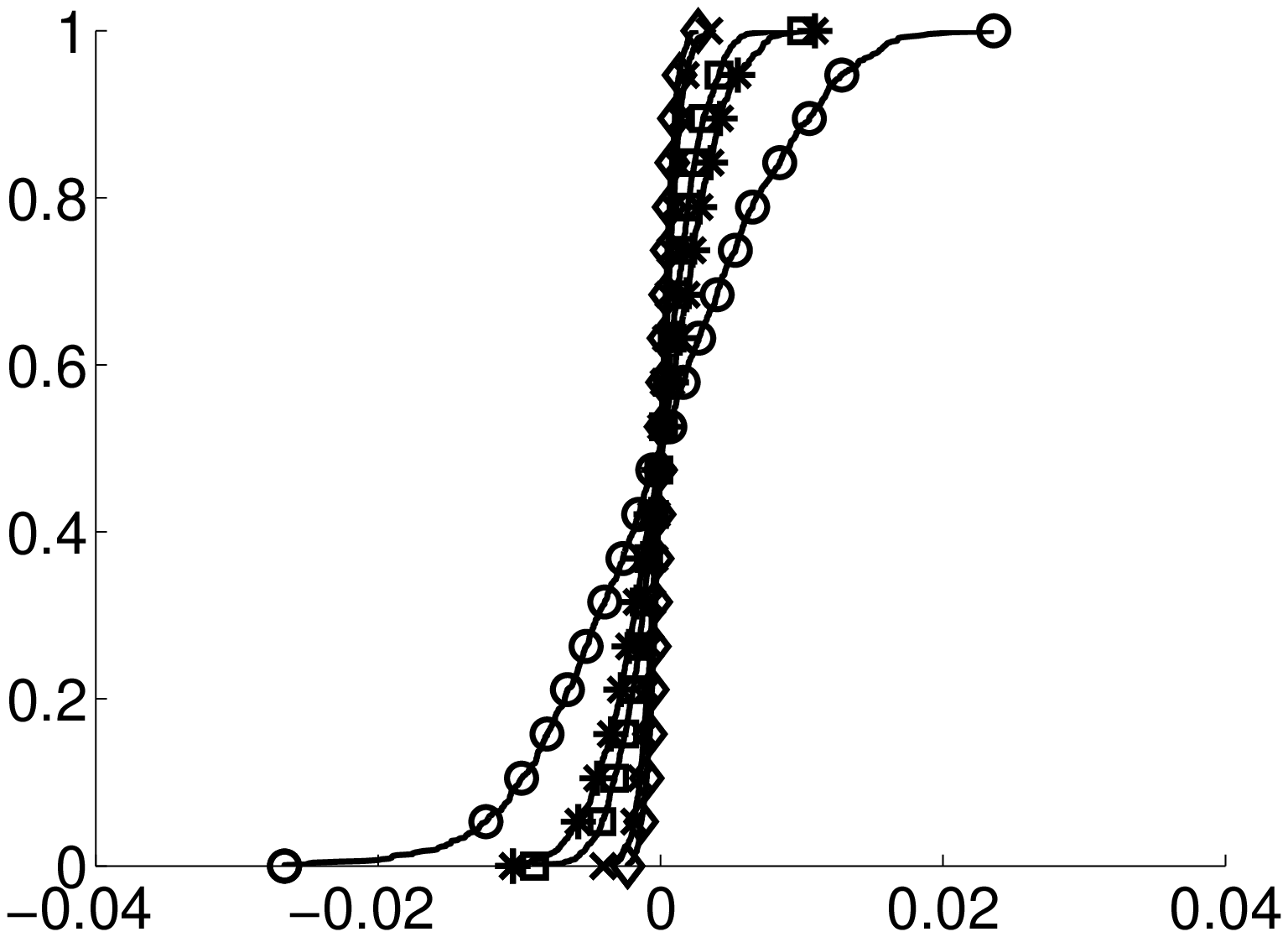}
		\caption{In-In}
		\label{sfig:normal_corr_inin}
	\end{subfigure}
	\caption{Plots of the empirical cumulative distribution of $\rho_\alpha^\beta$ for all four degree-degree 
	dependency types for ECM graphs  of different sizes with $\gamma_\pm = 2.1$. Each plot is based
	on $10^3$ realizations of the model.}
	\label{fig:normal_correlations_example}%
\end{figure}

\section{The directed Erased Configuration Model}

The directed Configuration Model (CM) starts with degree sequences $(D_i^+, D_i^-)_{1\le i \le N}$
that satisfy, for some $\mu > 0$,

\begin{subequations}
\begin{eqnarray}
	E = \sum_{i = 1}^N D_i^\pm &\sim& \mu N \\
	\sum_{i = 1}^N D_i^+ D_i^- &\sim& \mu^2 N\\
	\sum_{i = 1}^N (D_i^\pm)^p &\sim& N^{p/\gamma_\pm}, \quad p > \gamma_\pm.
\end{eqnarray}
\label{eq:configuration_model_scaling}
\end{subequations}

The stubs are then paired at random to form edges. This will in general constitute a graph with 
self-loops and multiple edges between nodes. If the degree variance is finite, then the 
probability of generating a simple graph is bounded away from zero and thus, by repeating the 
pairing step until such a graph is generated, we get a network randomly sampled from all networks 
of given size and degree sequences. This is called the Repeated Configuration Model (RCM). 

When the variance of the degrees is infinite, the probability of generating a simple graph converges 
to zero as the graph size increases, and therefore we need to enforce that the resulting graph is 
simple. For this we use the Erased Configuration Model (ECM), where, during the pairing, a new edge 
is removed if it already exists or if it is a self loop. Although this seems to be a strong 
alteration of the initial degree sequence, asymptotically, the degrees of the resulting network still 
follow the same distribution, see~\cite{chen2013}. For illustration, in Figure
\ref{fig:distributions_cm_ecm}, we plotted the degree distributions of and ECM graph of size $10^6$ 
before and after the removing of edges. Clearly there is hardly any difference between the two 
distributions. In particular the degree sequences of ECM graphs still satisfy 
(\ref{eq:configuration_model_scaling}). Unlike many other methods, random pairing of the stubs can be 
implemented very efficiently for even billions of nodes. Moreover, the ECM is computationally less 
expensive than the RCM, since we do not need to repeat the pairing. Therefore we suggest to use the ECM 
as a standard null-model. In the rest of the paper we will characterize the structural dependencies 
in the ECM.

%% file: degree_dependencies_ecm.tex
\begin{figure}[t]%
	\begin{subfigure}{.5\linewidth}
		\includegraphics[scale=0.35]{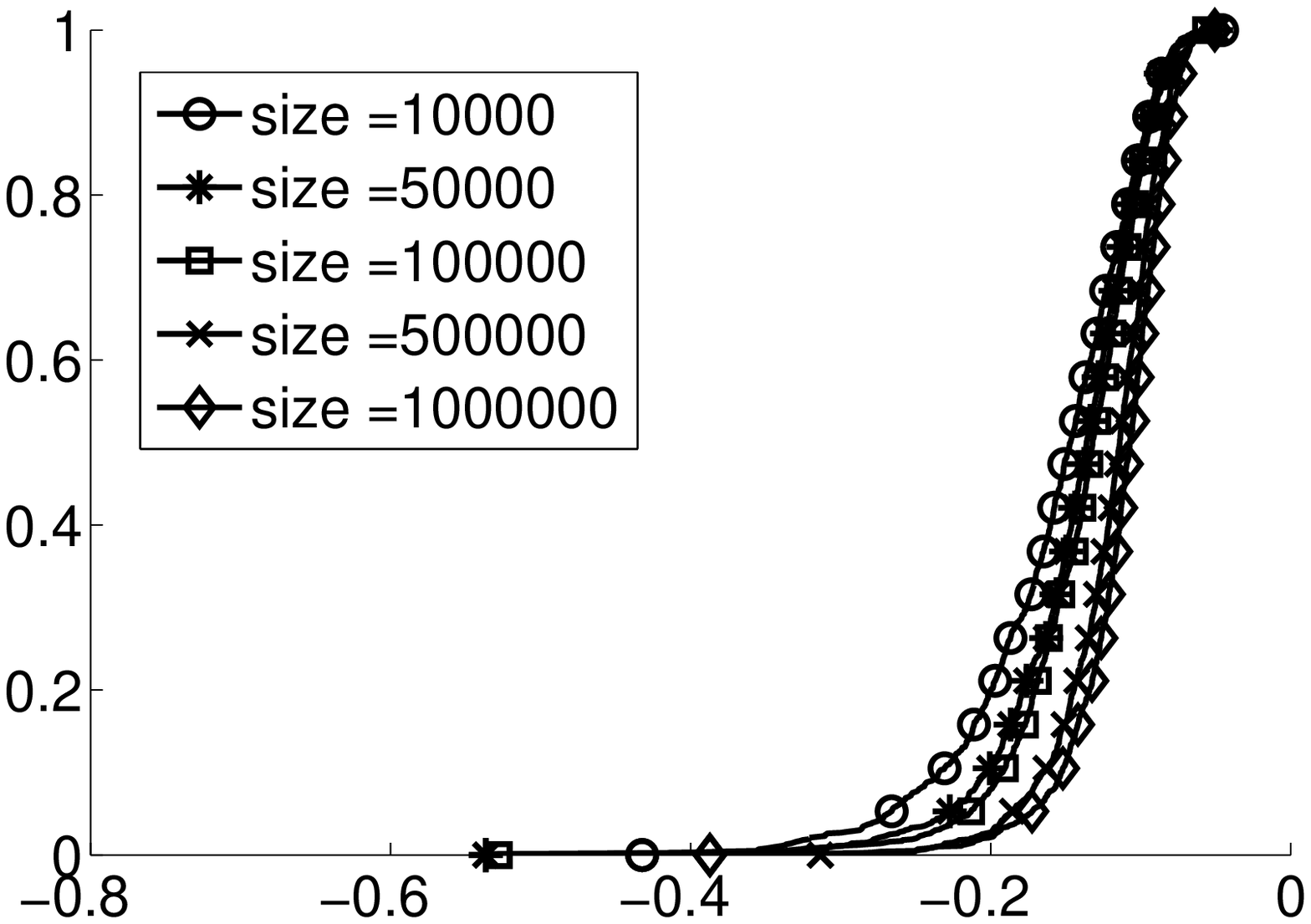}
		\caption{Out-In}
		\label{sfig:neg_corr_outin}
	\end{subfigure}~
	\begin{subfigure}{.5\linewidth}
		\includegraphics[scale=0.35]{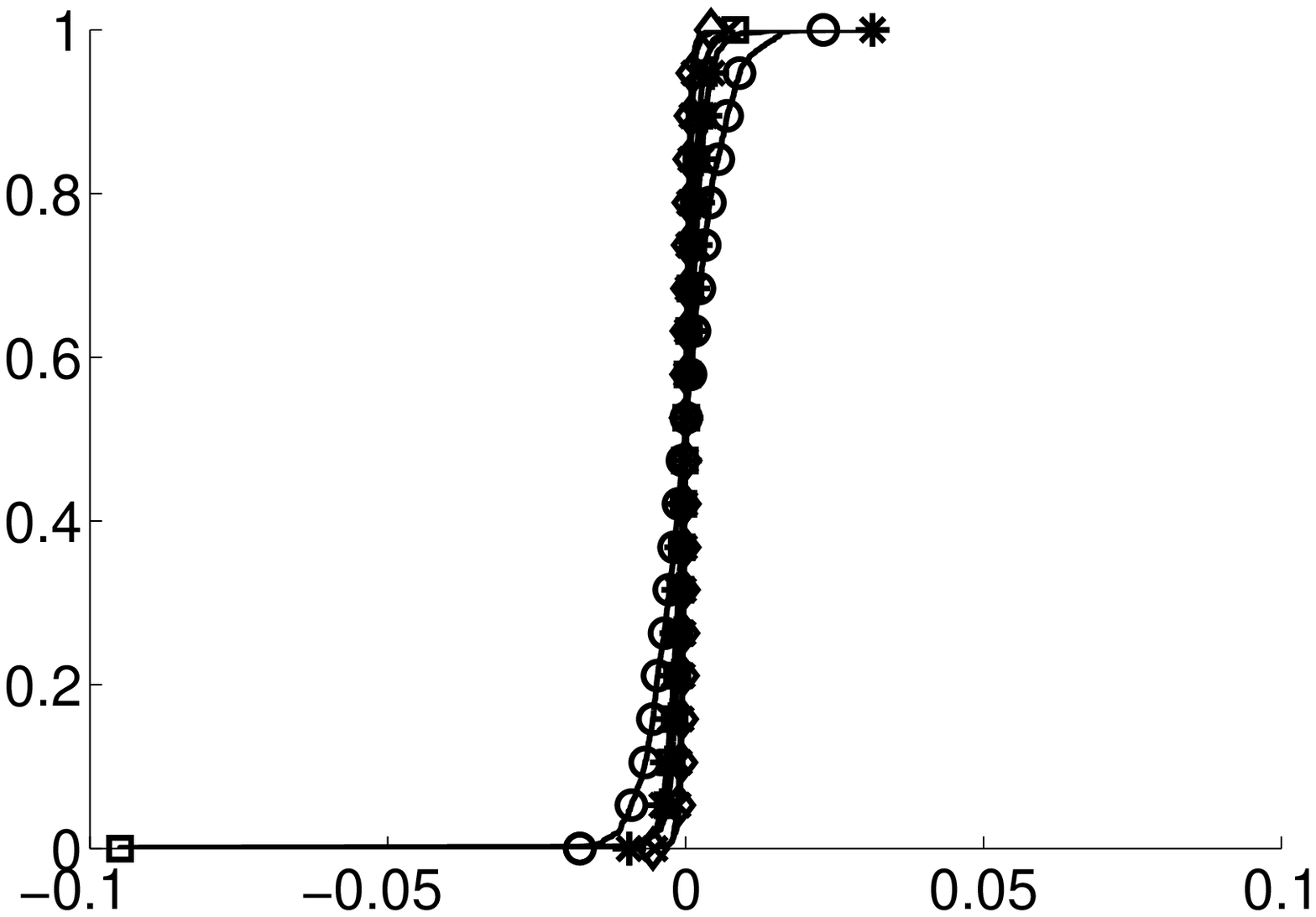}
		\caption{In-Out}
		\label{sfig:neg_correlations_inout}
	\end{subfigure}
	
	\begin{subfigure}{.5\linewidth}
		\includegraphics[scale=0.35]{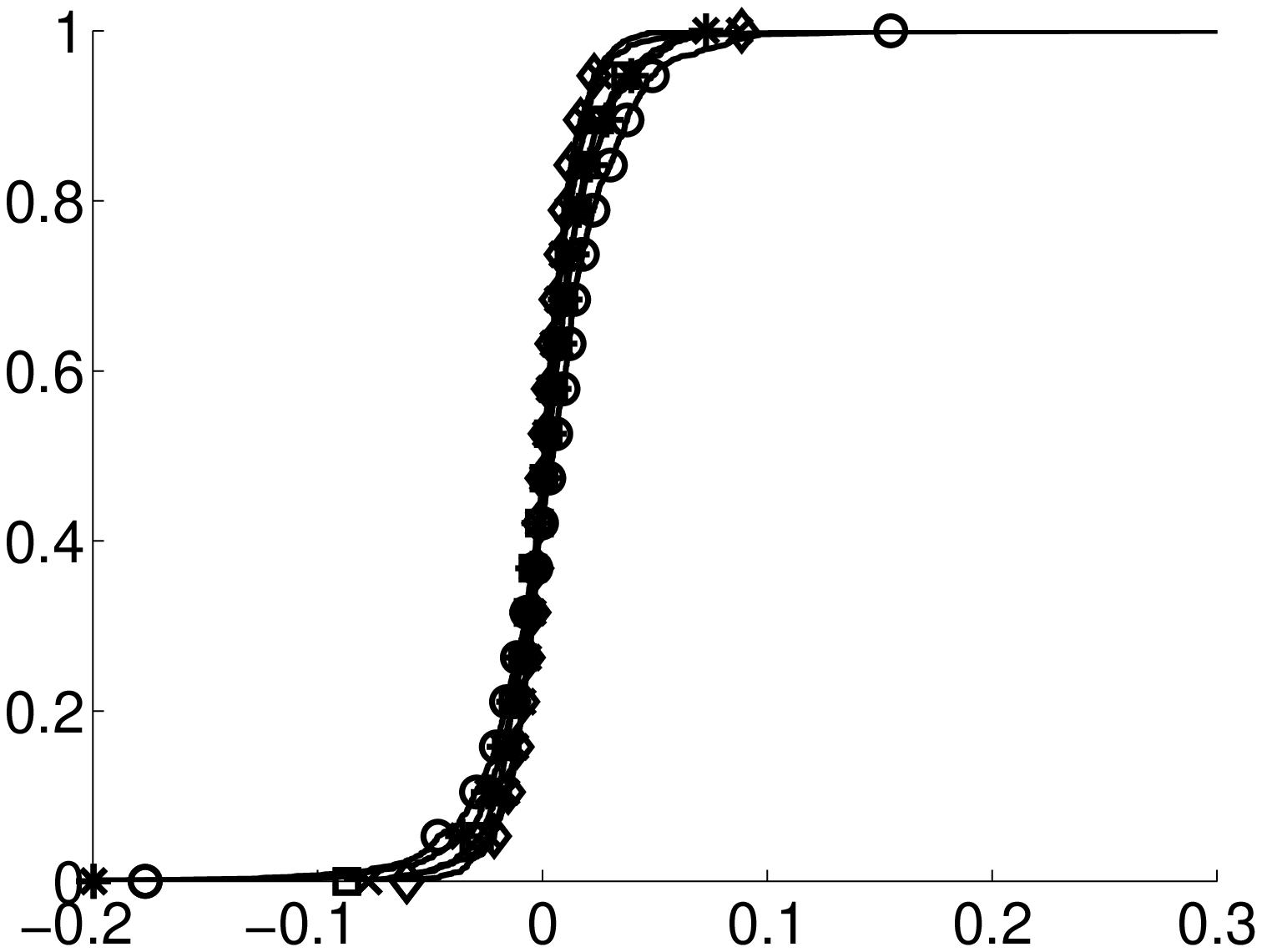}
		\caption{Out-Out}
		\label{sfig:neg_correlations_outout}
	\end{subfigure}~
	\begin{subfigure}{.5\linewidth}
		\includegraphics[scale=0.35]{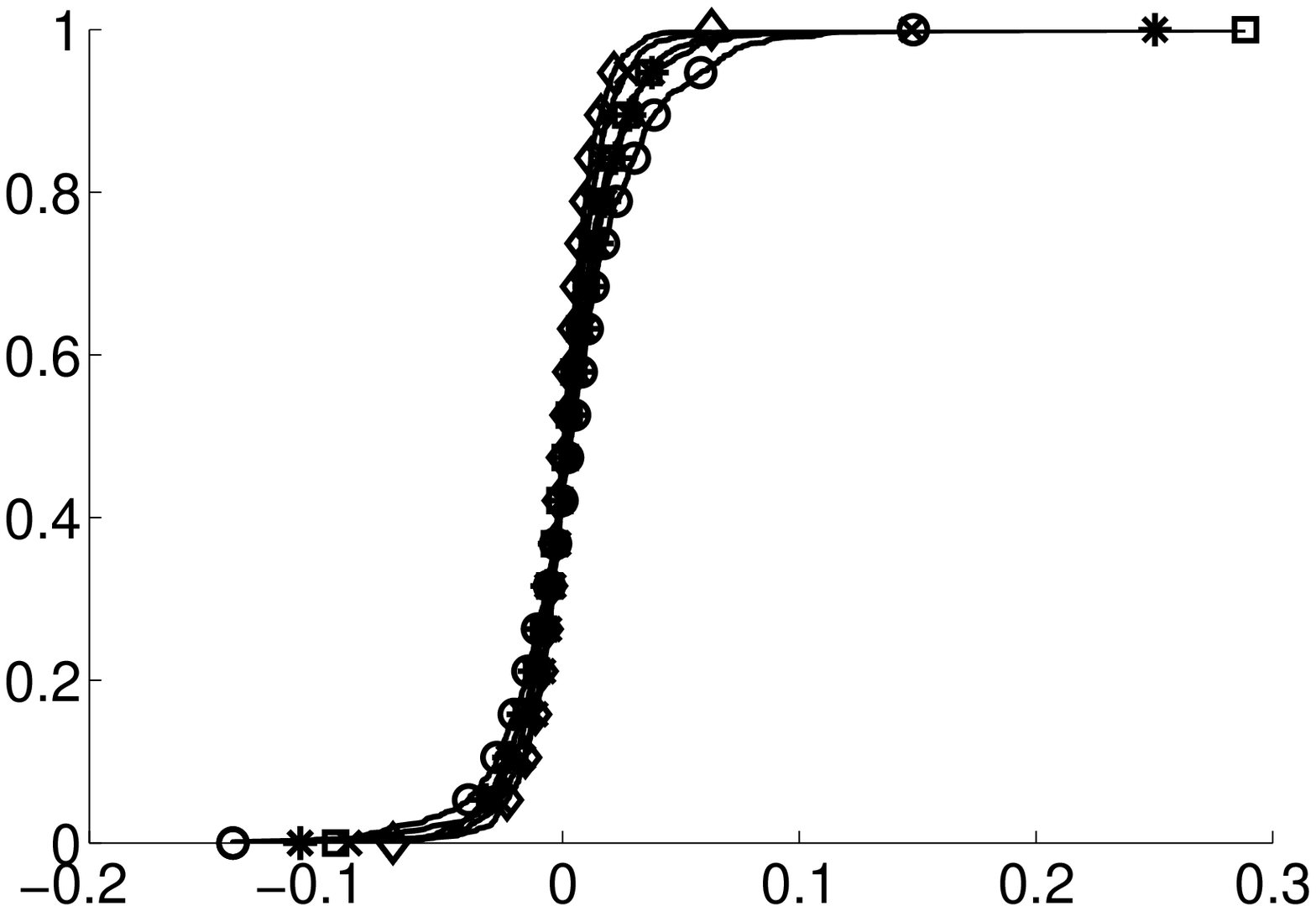}
		\caption{In-In}
		\label{sfig:neg_correlations_inin}
	\end{subfigure}
	\caption{Plots of the empirical cumulative distribution of $\rho_\alpha^\beta$ for all four degree-degree 
	dependency types for ECM graphs  of different sizes with $\gamma_\pm = 1.2$. Each plot is based
	on $10^3$ realizations of the model.}%
	\label{fig:negative_correlations_example}
\end{figure}

\section{Degree-degree dependencies in the ECM}\label{sec:dependencies_ecm}

It is clear that when we use the CM, i.e. allow for multiple edges and self loops, then our graphs
will have neutral mixing since all stubs are connected completely at random. For the ECM however, we
remove edges to make the graph simple, which has been shown~\cite{Maslov2004,Catanzaro2005} to give 
rise to negative correlations. Nevertheless, the ECM has asymptotically neutral mixing, which can be 
shown as follows.

Let $E_{ij}$ be the matrix counting the number of edges between $i$ and $j$ after the pairing and let 
$E_{ij}^c$ denote the matrix counting the number of removed edges between $i$ and $j$ by the ECM. 
Then for the CM it holds that $D_i^+ = \sum_{j = 1}^N E_{ij}$ while for the ECM we have $D_i^{+\,
\prime} = \sum_{j = 1}^N (E_{ij} - E_{ij}^c)$. Therefore, the difference between the empirical 
distributions of $D^\alpha_i$ and $D^\beta_j$, for an edge $i \to j$ sampled at random, in the CM and 
ECM, will be of the order $\sum_{i, j = 1}^N E_{ij}^c/E$, whose average, with respect to the degree 
sequences, converges to zero~\cite{Hoorn2014}, 

\begin{equation}
	\lim_{N \to \infty} \frac{1}{N} \sum_{i, j = 1}^N \la E_{ij}^c \ra = 0.
	\label{eq:convergence_erasededges}
\end{equation}
This implies that the values of $\rho_\alpha^\beta$ for an ECM graph will converge to that of a
CM graph, hence, asymptotically, $\rho_\alpha^\beta = 0$ and also $\overline{\rho}_\alpha^\beta = 
0 = \tau_\alpha^\beta$, for the ECM.

However, for finite realizations in the infinite variance regime, negative correlations are still 
observed. To illustrate this we plotted the empirical cumulative distribution functions of $\rho
_\alpha^\beta$ for graphs generated by the ECM with both finite and infinite degree variance, see 
Figure~\ref{fig:normal_correlations_example} and Figure~\ref{fig:negative_correlations_example}, 
respectively. In addition, Table~\ref{tbl:outin_averages} contains the average values for all four
correlation types in the infinite variance regime. One immediately observes that the Out-In 
dependency in ECM graphs with infinite variance, Figure~\ref{sfig:neg_corr_outin}, displays strong 
structural negative correlations which decrease as the network grows, while for the other three 
dependency types the values are concentrated around zero. Moreover, we see, Figure
\ref{fig:normal_correlations_example}, that all four dependency types behave similar when the 
variance of the degrees is finite.

\input{tables_spearman_outin_averages}

These negative Out-In correlations ($\rho_+^-$) can be explained by first observing that multiple 
edges are more likely to start in a node of large out-degree and end in a node of large in-degree,
since these are more likely to be sampled. 
Now, consider the algorithm as first connecting all stubs at random and then removing self loops 
and merging multiple edges. By construction, immediately after the pairing the network will have 
neutral mixing. When merging multiple edges we will often delete connections from nodes of large 
out-degree to nodes of large in-degree. Such edges have contributed positively into $\rho_{+}^{-}$, 
thus, deleting them will shift $\rho_{+}^{-}$ from zero in the CM to a negative value in the ECM. The 
other three dependency types are not effected since the out- and in-degree of a node in the ECM are 
independent. 

Motivated by the analysis in this section, we will further focus on the behavior of $\rho_{+}^{-}$ in 
the infinite-variance case, $1< \gamma_+,\gamma_-\le 2$, as the only scenario where we observe prominent 
structural correlations. We will discuss other scenarios in Section~\ref{sec:other_cases}.

%% file: tables_spearman_outin_averages.tex
\begin{table}%
\centering
\begin{tabular}{c|cccc}
	$N$ & $\langle \rho_+^- \rangle$ & $\langle \rho_-^+ \rangle$ & $\langle \rho_+^+ \rangle$ 
		& $\langle \rho_-^- \rangle$ \\
	\hline
	10000 	& -0.1568	& -0.0001	&	0.0039	&	0.0048 \\
	50000		&	-0.1439 &	0.0001	&	0.0014	&	0.0029 \\
	100000	&	-0.1388 &	-0.0001	&	0.0026	&	0.0028 \\
	500000	&	-0.1198	& 0.0001	&	0.0011 	&	0.0017 \\
	1000000	&	-0.1131	& 0.0000	&	0.0009	&	0.0002
\end{tabular}
\caption{The average values for $\rho$ for all four degree-degree dependencies types, for ECM 
graphs of different sizes, with $\gamma_\pm = 1.2$, based on $10^3$ realizations of the model.}
\label{tbl:outin_averages}
\end{table}

%% file: analysis_correlations.tex
\section{Scaling of the Out-In degree-degree dependency in the ECM}

We will determine the scaling of $\rho_+^-$ as a function of the exponents $\gamma_\pm$. That is, 
we will find coefficients $f(\gamma_+, \gamma_-)$ such that 
\[
	\frac{\rho_+^- - \la \rho_+^- \ra}{N^{f(\gamma_+, \gamma_-)}}
\] 
converges to some limiting distribution. Here the expectation $\la \rho_+^- \ra$ is taken over 
all possible graphs of size $N$, generated by the ECM, with degree sequences satisfying 
(\ref{eq:configuration_model_scaling}). We note that although $\la \rho_+^- \ra$ is of
similar order as the typical spreading of $\rho_+^-$, the latter, which we 
are going to evaluate, will define the magnitude of the structural negative correlations. 

We obtain the scaling exponents $f(\gamma_+, \gamma_-)$ by establishing upper bounds on the scaling, 
and then show empirically that these bounds are tight. The scaling is an important quantity, 
characterizing the spread around the sample mean of $\rho_+^-$ as a function of $N$. Roughly, 
this tells us how much the measured values on a ECM graph of size $N$ can deviate from the average 
and therefore enable us to asses the significance of the measured correlations of the corresponding 
real world networks. 

\subsection{Scaling of the erased number of edges}

As we discussed in the previous section, the structural negative correlations appear after multiple 
edges and self-loops are erased. Hence, part of the scaling of $\rho_+^-$ comes from the scaling of the 
average total number of erased edges. The latter scaling has a phase transition, which we will show 
by establishing two different upper bounds.
 
For the first upper bound, observe that
\begin{equation}
	\sum_{i,j = 1}^N E_{ij}^c =  \sum_{i = 1}^N S_{ii} + \sum_{i,j = 1}^N M_{ij},
\label{eq:erased_edges_bound_SM}
\end{equation}
where $S$ is the diagonal matrix counting the number of self loops and $M$ is the zero diagonal
matrix that counts the excess edges, so $M_{ij} = k > 0$ means that $E_{ij} = k + 1$. For the self 
loops it holds that
\begin{equation}
	\la S_{ii} \ra =  \frac{D_i^+ D_i^-}{E}.
\label{eq:self_loops}
\end{equation}
If we now take the total number of pairs of edges between $i$ and $j$ as an upper bound for 
$M_{ij}$, then
\begin{equation}
	\la M_{ij} \ra \le \frac{(D_i^+)^2(D_j^-)^2}{E^2}.
\label{eq:multiple_edges}
\end{equation}
Applying~(\ref{eq:self_loops}) and~(\ref{eq:multiple_edges}) to~(\ref{eq:erased_edges_bound_SM}) we
get
\begin{equation}
	\sum_{i, j = 1}^N\frac{\la E^c_{ij} \ra}{E} \le \frac{\sum_{i, j = 1}^N (D_i^+)^2(D_j^-)^2}{E^3} + 
	\frac{\sum_{i = 1}^N D_i^+ D_i^-}{E^2}.
	\label{eq:erased_edges_bound1}
\end{equation}
We remark that if the second moment of both the out- and in-degree exists, then this upper bound 
scales as $N^{-1}$. When this is not the case, we get the scaling from
(\ref{eq:configuration_model_scaling}) as
\begin{equation}
	\frac{1}{E} \sum_{i,j = 1}^N \la E^c_{ij} \ra = O\left(N^{(2/\gamma_+) + (2/\gamma_-) - 3}\right).
	\label{eq:first_scaling_bound}
\end{equation}

The upper bound~(\ref{eq:first_scaling_bound}) is rather crude in the sense that for certain $1 < 
\gamma_\pm \le 2$, we have $(2/\gamma_+) + (2/\gamma_-) > 3$ so that the right-hand side of 
(\ref{eq:first_scaling_bound}) becomes infinite as $N\to\infty$. 

To get a more precise upper bound let $p(n,m,L)$ denote the probability that none of the outbound stubs 
from a set of size $n$ connect to an inbound stub from a set of size $m$, given that the total number 
of available stubs is $L$. We will establish a recursive relation for $p(D_i^+, D_j^-, E)$ by adopting 
the analysis from~\cite{Hofstad2005}, Section 4. Similarly we get, by conditioning on whether we pick 
an inbound stub of $i$ or not,
\begin{align*}
	p(D_i^+, D_j^-, E) \le \left(1 - \frac{D_j^-}{E}\right) p(D_i^+ - 1, D_j^-, E - 1),
\end{align*}
where the upper bound comes from neglecting the event $D_i^+ + D_j^- > E$, in which case
$p(D_i^+, D_j^-, E) = 0$. Continuing the recursion yields
\begin{equation*}
	p(D_i^+, D_j^-, E) \le \prod_{k = 0}^{D_i^+ - 1} \left(1 - \frac{D_j^-}{E - 
	k}\right),
\end{equation*}
and a first order Taylor expansion then gives 
\begin{equation}
	p(D_i^+, D_j^-, E) \le e^{-D_i^+ D_j^-/E}.
	\label{eq:prob_no_connection_ij}
\end{equation}
Now, recall that $E_{ij}$ denotes the total number of edges between $i$ and $j$ in the CM, before the 
removal step. Therefore,
\begin{equation*}
	\la E_{ij}^c \ra = \la E_{ij} \ra - (1 - p(D_i^+, D_j^-, E)).
\end{equation*}
Since $E = \sum_{i,j = 1}^N \la E_{ij} \ra$ it follows that
\begin{equation}
	\frac{1}{E} \sum_{i, j = 1}^N \la E_{ij}^c \ra = 1 - \frac{N^2}{E} + \frac{1}{E} \sum_{i, j = 1}^N 
	p(D_i^+, D_j^-, E)
\label{eq:average_erased_edges}
\end{equation}

Hence, by plugging~(\ref{eq:prob_no_connection_ij}) into~(\ref{eq:average_erased_edges}) we arrive 
at the following upper bound for the total average number of erased edges,

\begin{equation}
	\frac{1}{E} \sum_{i, j = 1}^N \la E_{ij}^c \ra \le 1 - \frac{N^2}{E} + \frac{1}{E} \sum_{i, j = 1}^N 
	e^{-D_i^+ D_j^-/E}.
\label{eq:erased_edges_bound2}
\end{equation}

The right hand side of~(\ref{eq:erased_edges_bound2}) can be slightly rewritten to obtain a more 
informative expression, which is the product of $N^2/E$ and the term
\begin{equation}
	\frac{1}{E} \sum_{i,j = 1}^N \frac{D^+_i D^-_j}{N^2} - 1 + \sum_{i,j = 1}^N \frac{e^{-(D^+_i 
	D^-_j)/E}}{N^2}.
\label{eq:empirical_tauberian_form}
\end{equation}
Next, we note that~(\ref{eq:empirical_tauberian_form}) can be seen as an empirical form of
\begin{equation}
\frac{1}{N\mu} \la \xi \ra - 1 + \la e^{-\xi/(N\mu)} \ra,
\label{eq:tauberian_form}
\end{equation}
where, letting $\gamma_{\text{min}} = \min\{\gamma_+, \gamma_-\}$, $\xi$ has distribution
\[
	P_{\xi}(k) \sim k^{-(\gamma_{\text{min}} + 1)},
\]
and $\la \xi \ra = \mu^2$. From a classical Tauberian Theorem for regularly varying random 
variables, see for instance~\cite{Bingham1974} Theorem A, it follows that
(\ref{eq:tauberian_form}) scales as $N^{-\gamma_{\text{min}}}$. When we replace $E$ by $\mu N$ in
(\ref{eq:empirical_tauberian_form}), we obtain
\begin{equation}
	\frac{1}{\mu N} \sum_{i,j = 1}^N \frac{D_i^+ D_j^-}{N^2} - 1 + \sum_{i,j = 1}^N 
	\frac{e^{-D_i^+ D_j^-/(\mu N)}}{N^2}
	\label{eq:empirical_tauberian_form_2}
\end{equation}
and observe that~(\ref{eq:tauberian_form}) is the expectation of 
(\ref{eq:empirical_tauberian_form_2}). The function $f(x) = x - 1 + e^{-x}$ is positive, hence, it
follows that~(\ref{eq:empirical_tauberian_form_2}) and~(\ref{eq:tauberian_form}) have the same 
scaling, $N^{-\gamma_{\text{min}}}$. Finally, the difference between 
(\ref{eq:empirical_tauberian_form}) and~(\ref{eq:empirical_tauberian_form_2}) is dominated by the 
term
\[
	\left|\frac{1}{E} - \frac{1}{\mu N}\right| = O\left(N^{-2}\left|E - \mu N\right|\right).
\]
Recall that $\sum_{i = 1}^n D_i^+ = E = \sum_{i = 1}^n D_i^-$. Hence, we obtain from the Central 
Limit Theorem for regularly varying random variables, see \cite{Whitt2002}, that 
\[
	N^{-2}|E - \mu N| = O\left(N^{-2 + 1/\gamma_{\text{min}}}\right).
\]
which dominates $N^{- \gamma_{\text{min}}}$ when $1 < \gamma_\pm \le 2$. Summarizing, we have
that~(\ref{eq:empirical_tauberian_form}) scales as $O(N^{-2 + 1/\gamma_{\text{min}}})$ and hence,
since $N^2/E = O(N)$, it follows that
\begin{equation}
	\frac{1}{E} \sum_{i, j = 1}^N \la E_{ij}^c \ra = O(N^{-1 + 1/\gamma_{\text{min}}}).
	\label{eq:second_scaling_bound}
\end{equation}

The scaling in~(\ref{eq:second_scaling_bound}) is related to that of the structural cut off described 
in~\cite{Boguna2004}, adjusted to the setting of directed networks with degree distributions 
(\ref{eq:in_out_distribution}). Moreover, comparing~(\ref{eq:second_scaling_bound}) to 
(\ref{eq:first_scaling_bound}) we observe a phase transition, with respect to the tail exponents 
$\gamma_\pm$ of the degree distributions, in the scaling of the average total  number of removed edges 
in the ECM, which will induce a phase transition in the scaling of the Out-In degree-degree dependency.

\begin{figure}[htp]%
	\centering
		\input{figures_scaling_scalingarea}
	\caption{Plot of the different scaling regimes for $\rho_+^-$. The scaling terms for each of the
		three regions can be found in Table~\ref{tbl:scaling_terms}.
		The Roman numerals indicate the three different choices of $\gamma_+$ and 
		$\gamma_-$, used in Figure~\ref{fig:phase_transitions} and~\ref{fig:phase_transitions_outout}, to 
		illustrated the	different regimes.}
	\label{fig:scaling_area}
\end{figure}
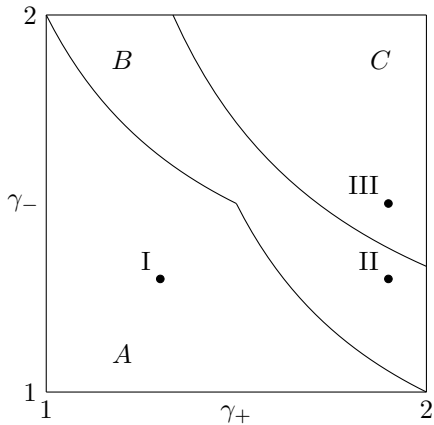

\begin{figure}[htp]%
	\begin{subfigure}{0.32\linewidth}
		\includegraphics[scale=0.26]{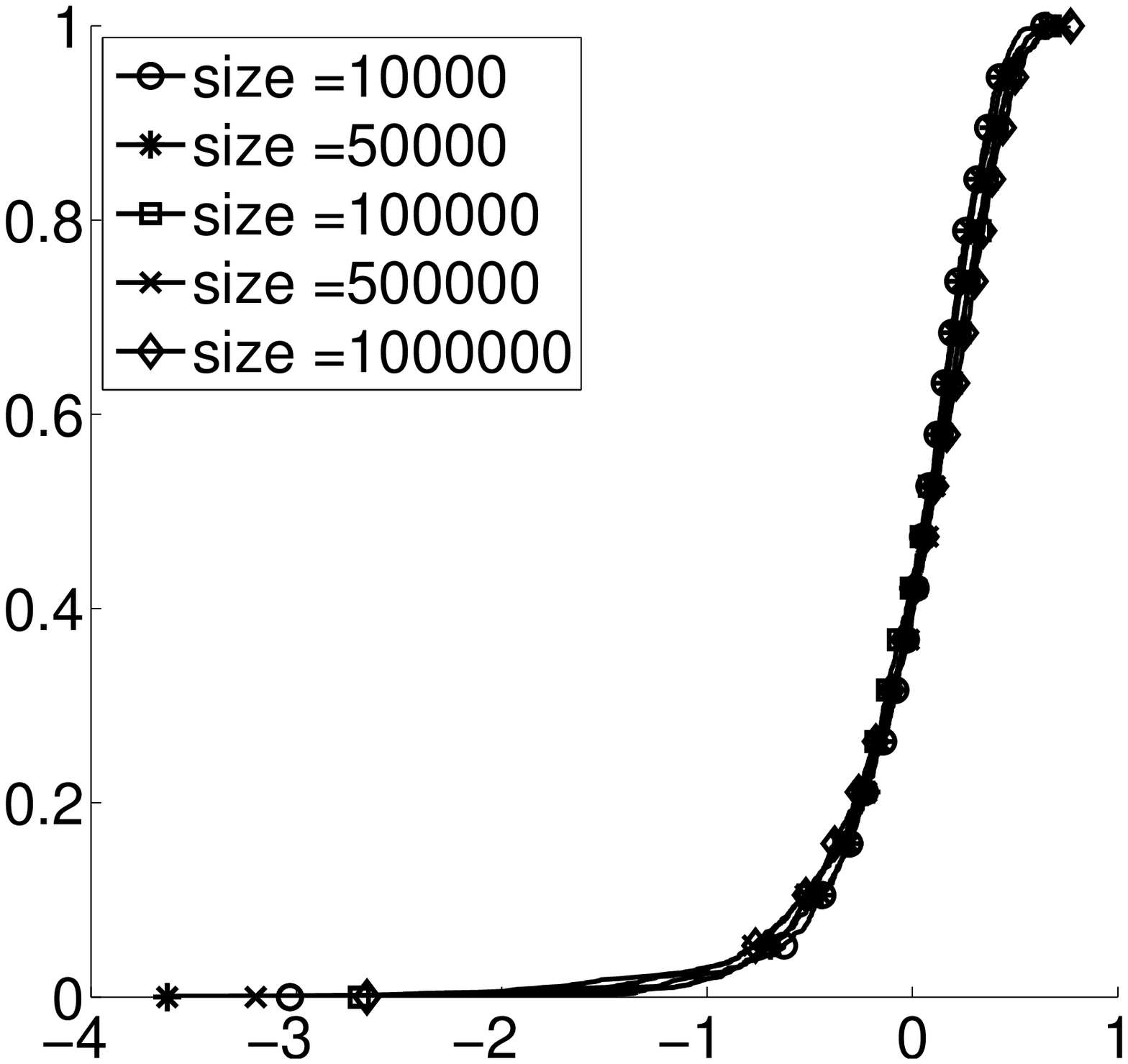}
		\caption{I - $N^{1/\gamma_{\text{min}} - 1}$}
		\vspace{5pt}
	\end{subfigure} 
	\begin{subfigure}{0.32\linewidth}
		\includegraphics[scale=0.26]{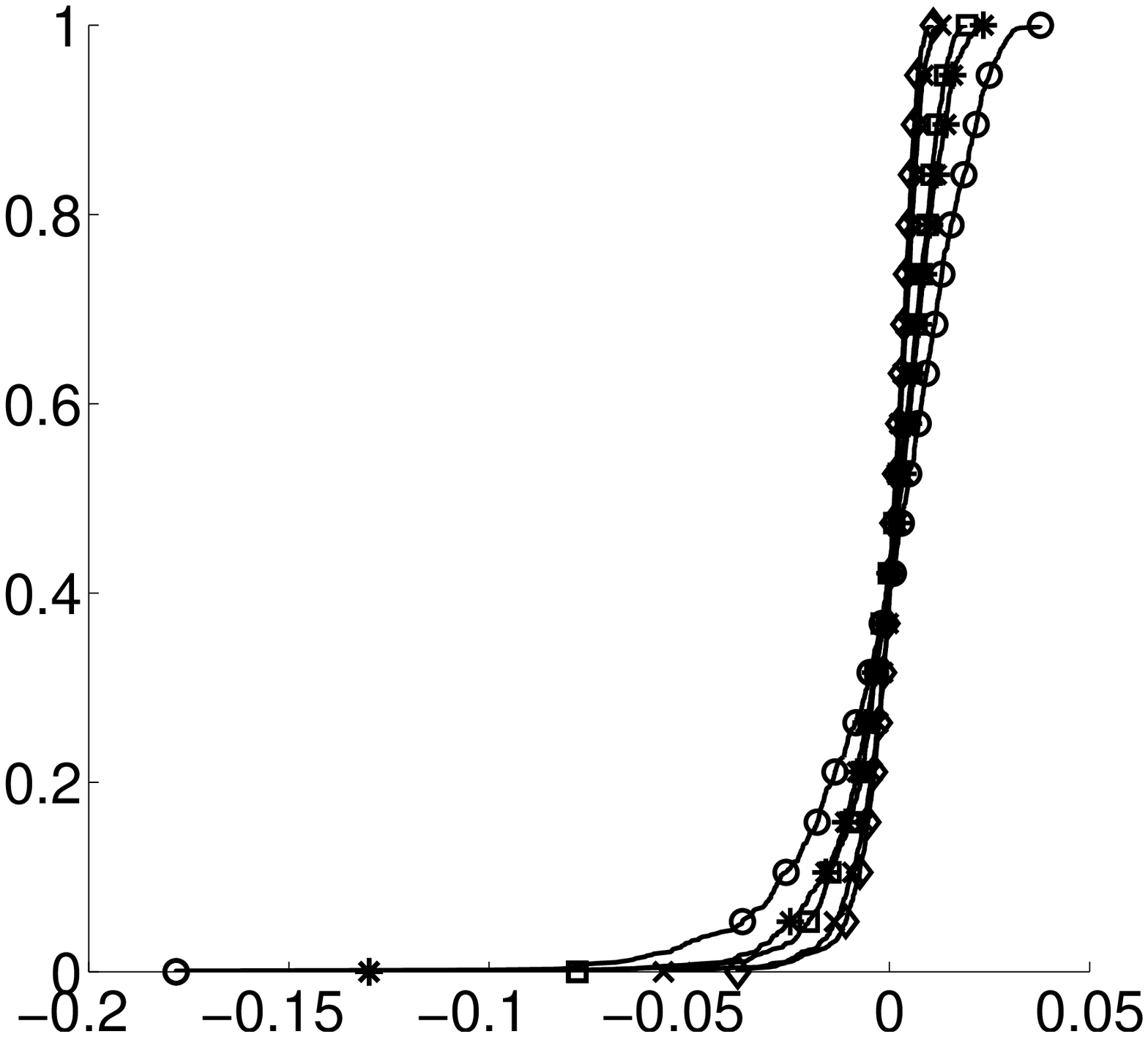}
		\caption{I - $N^{(2/\gamma_+) + (2/\gamma_-) - 3}$}
		\vspace{5pt}
	\end{subfigure}
	\begin{subfigure}{0.32\linewidth}
		\includegraphics[scale=0.26]{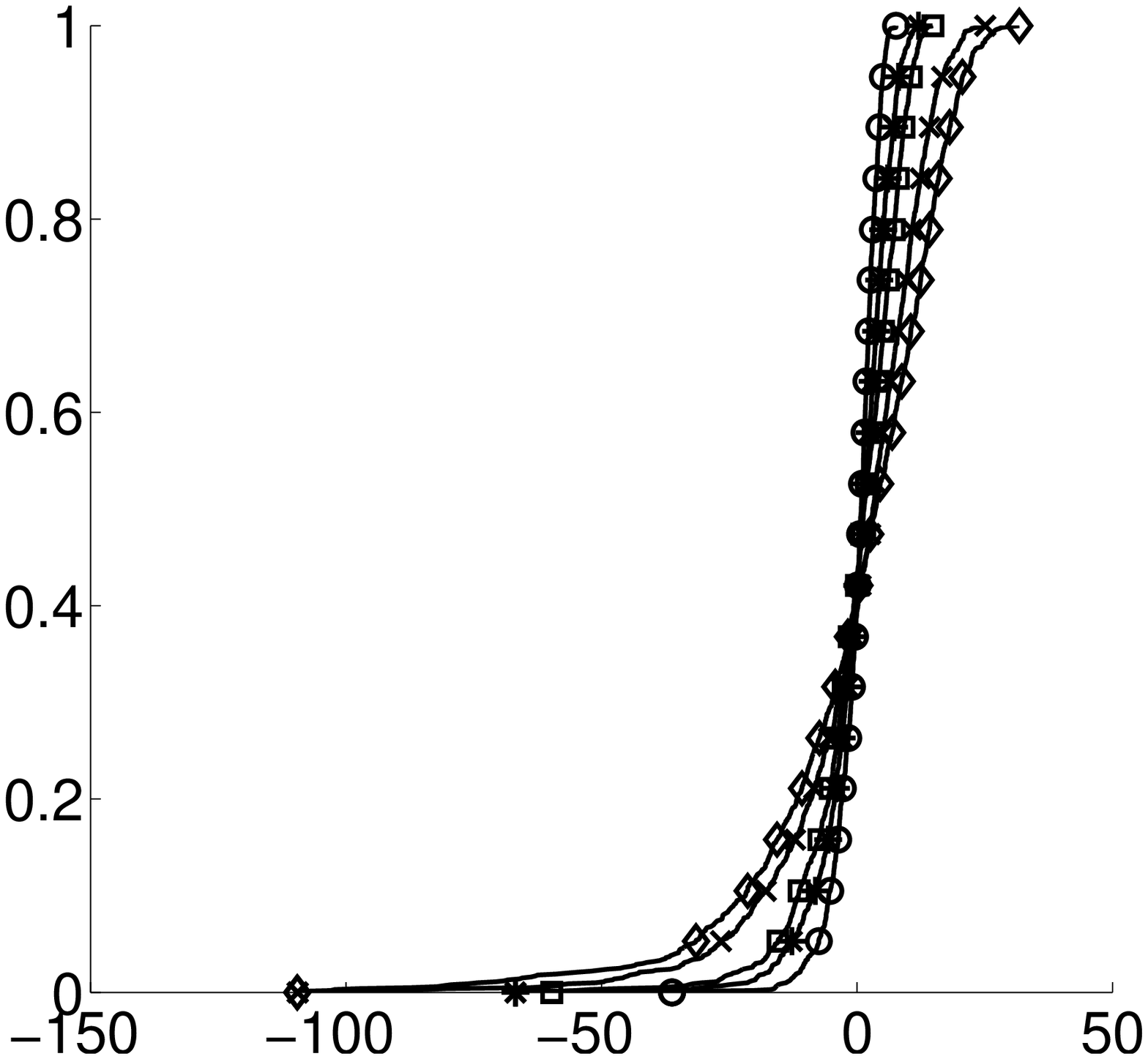}
		\caption{I - $N^{-1/2}$}
		\vspace{5pt}
	\end{subfigure}
	\begin{subfigure}{0.32\linewidth}
		\includegraphics[scale=0.26]{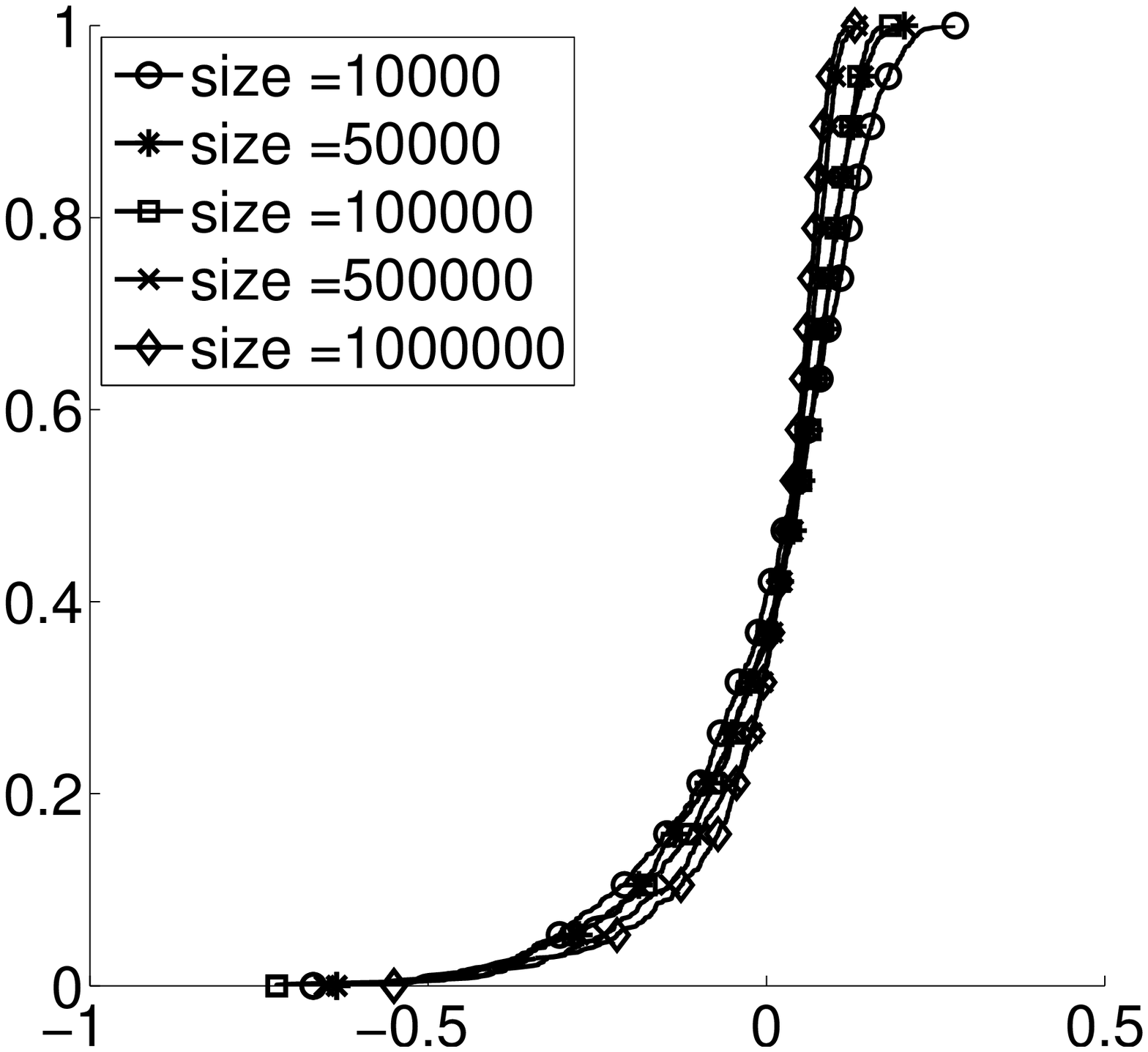}
		\caption{II - $N^{1/\gamma_{\text{min}} - 1}$}
		\vspace{5pt}
	\end{subfigure} 
	\begin{subfigure}{0.32\linewidth}
		\includegraphics[scale=0.26]{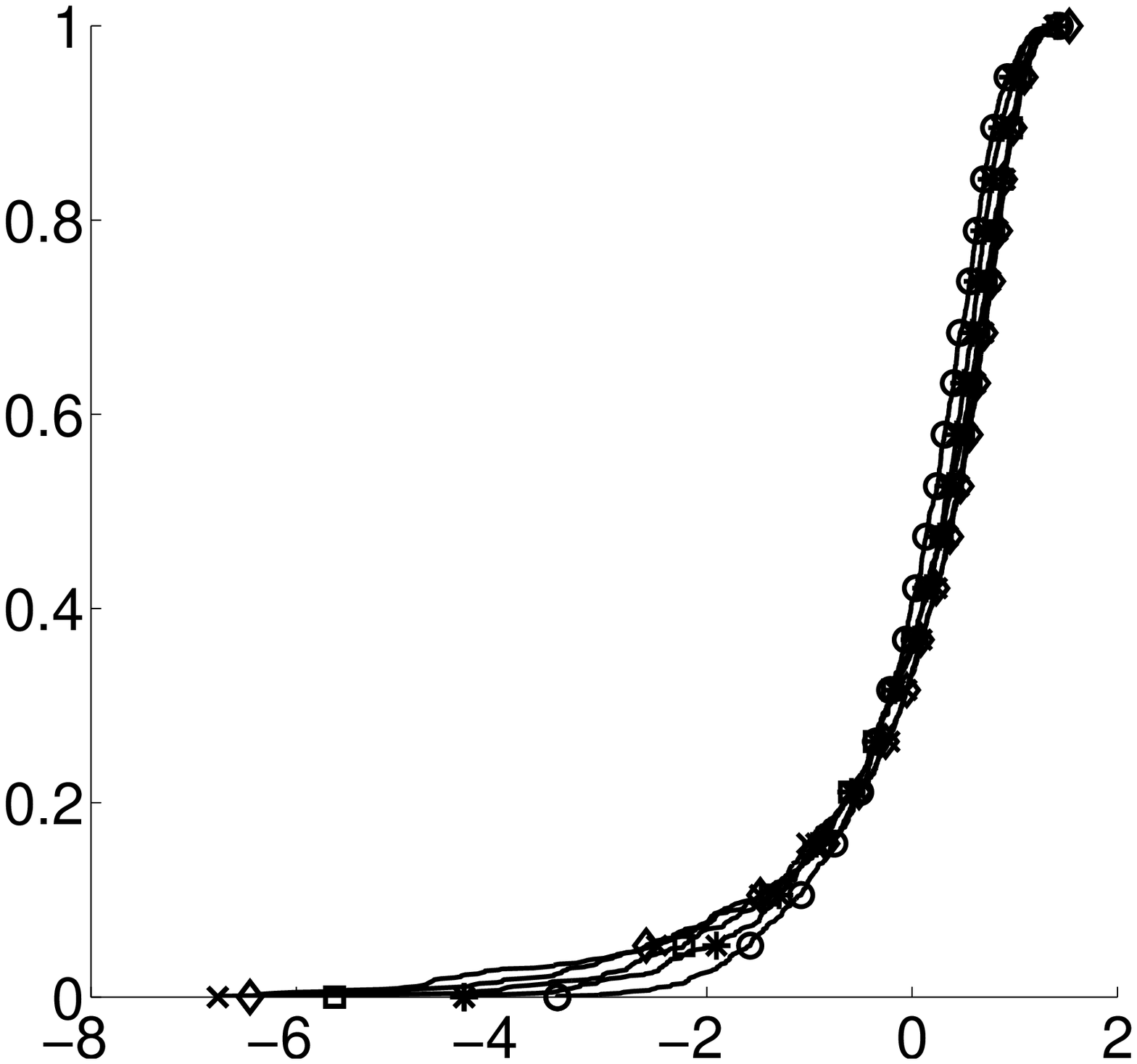}
		\caption{II - $N^{(2/\gamma_+) + (2/\gamma_-) - 3}$}
		\vspace{5pt}
	\end{subfigure}
	\begin{subfigure}{0.32\linewidth}
		\includegraphics[scale=0.26]{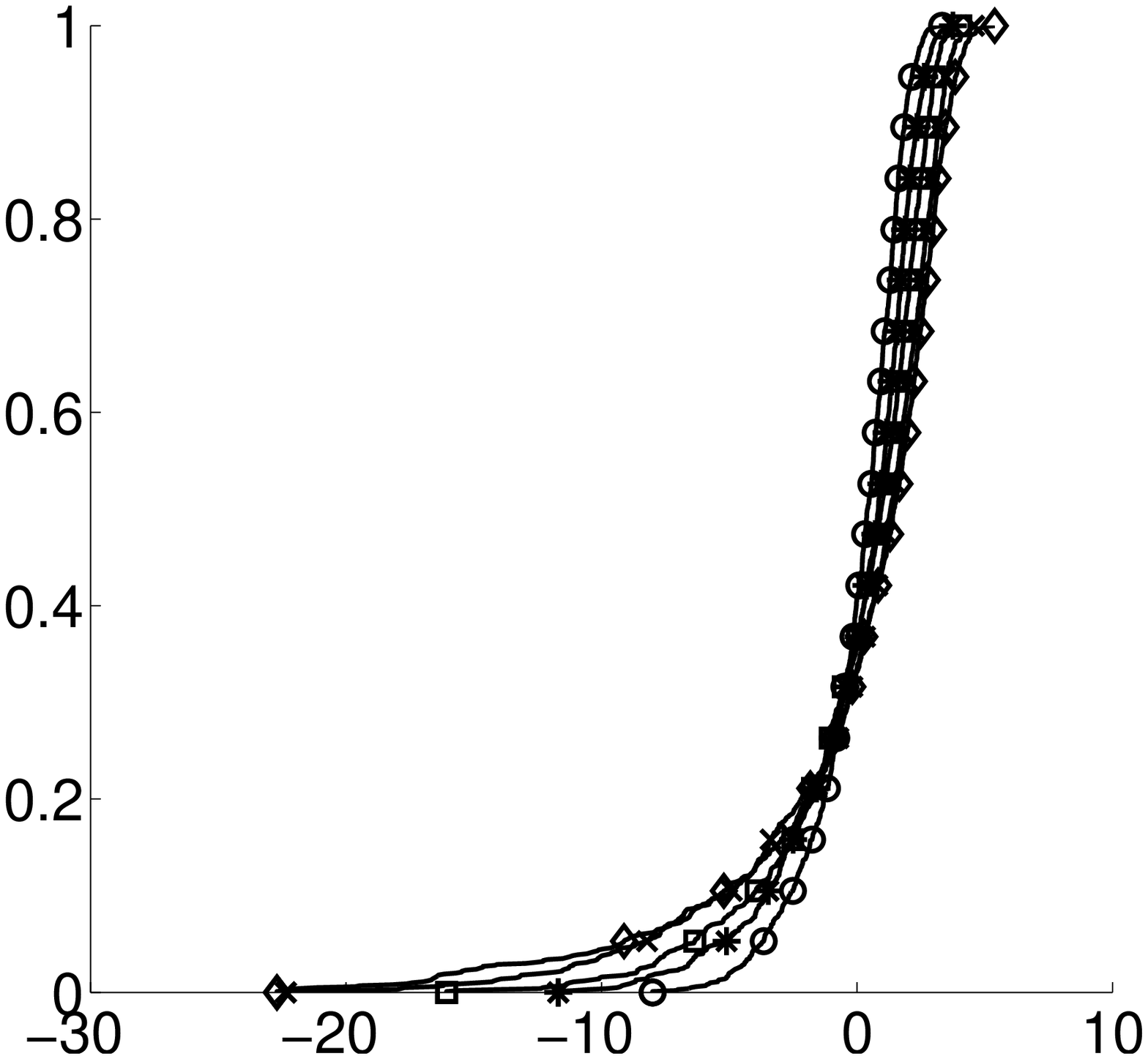}
		\caption{II - $N^{-1/2}$}
		\vspace{5pt}
	\end{subfigure}
	\begin{subfigure}{0.32\linewidth}
		\includegraphics[scale=0.26]{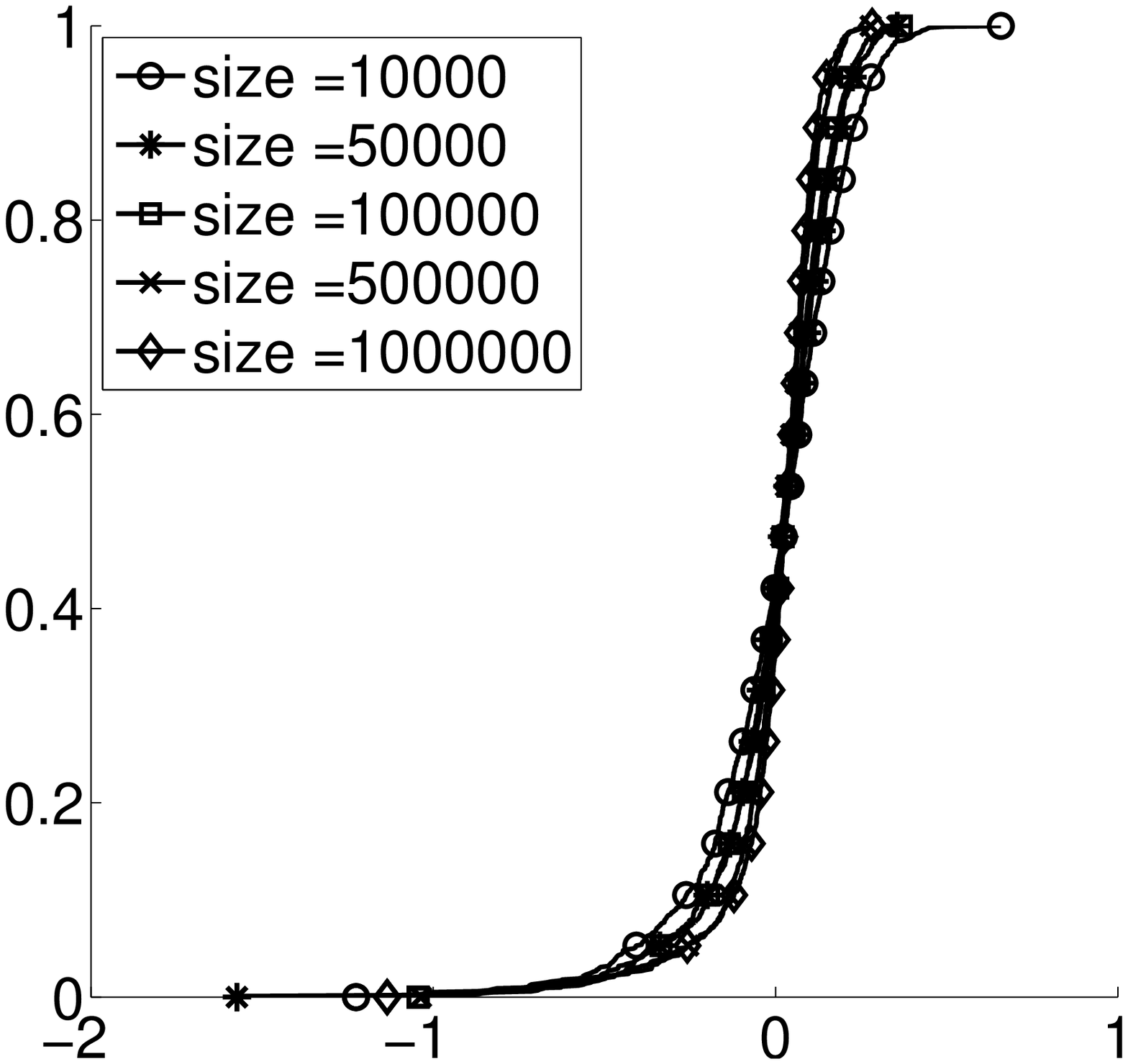}
		\caption{III - $N^{1/\gamma_{\text{min}} - 1}$}
	\end{subfigure} 
	\begin{subfigure}{0.32\linewidth}
		\includegraphics[scale=0.26]{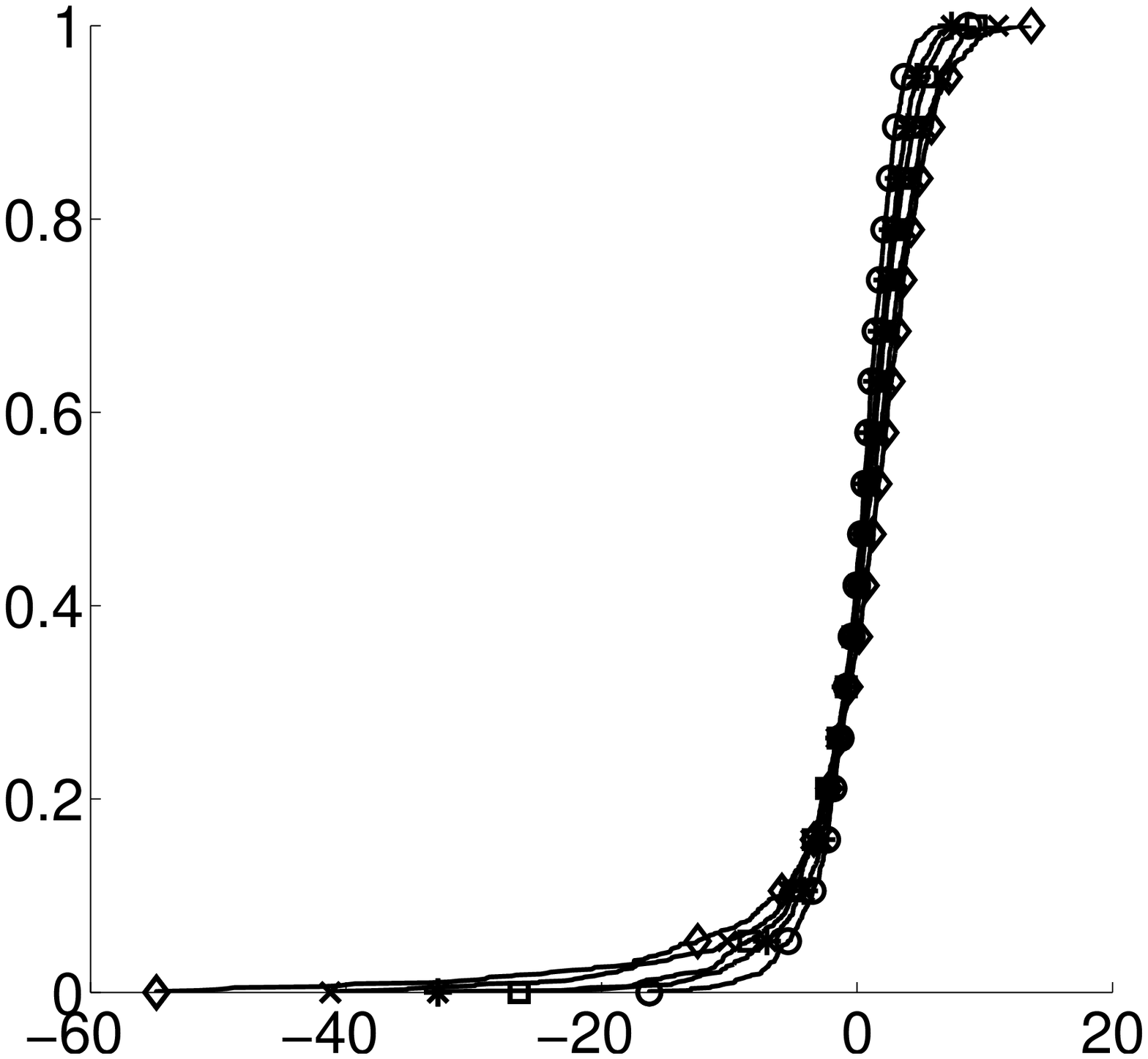}
		\caption{III - $N^{(2/\gamma_+) + (2/\gamma_-) - 3}$}
	\end{subfigure}
	\begin{subfigure}{0.32\linewidth}
		\includegraphics[scale=0.26]{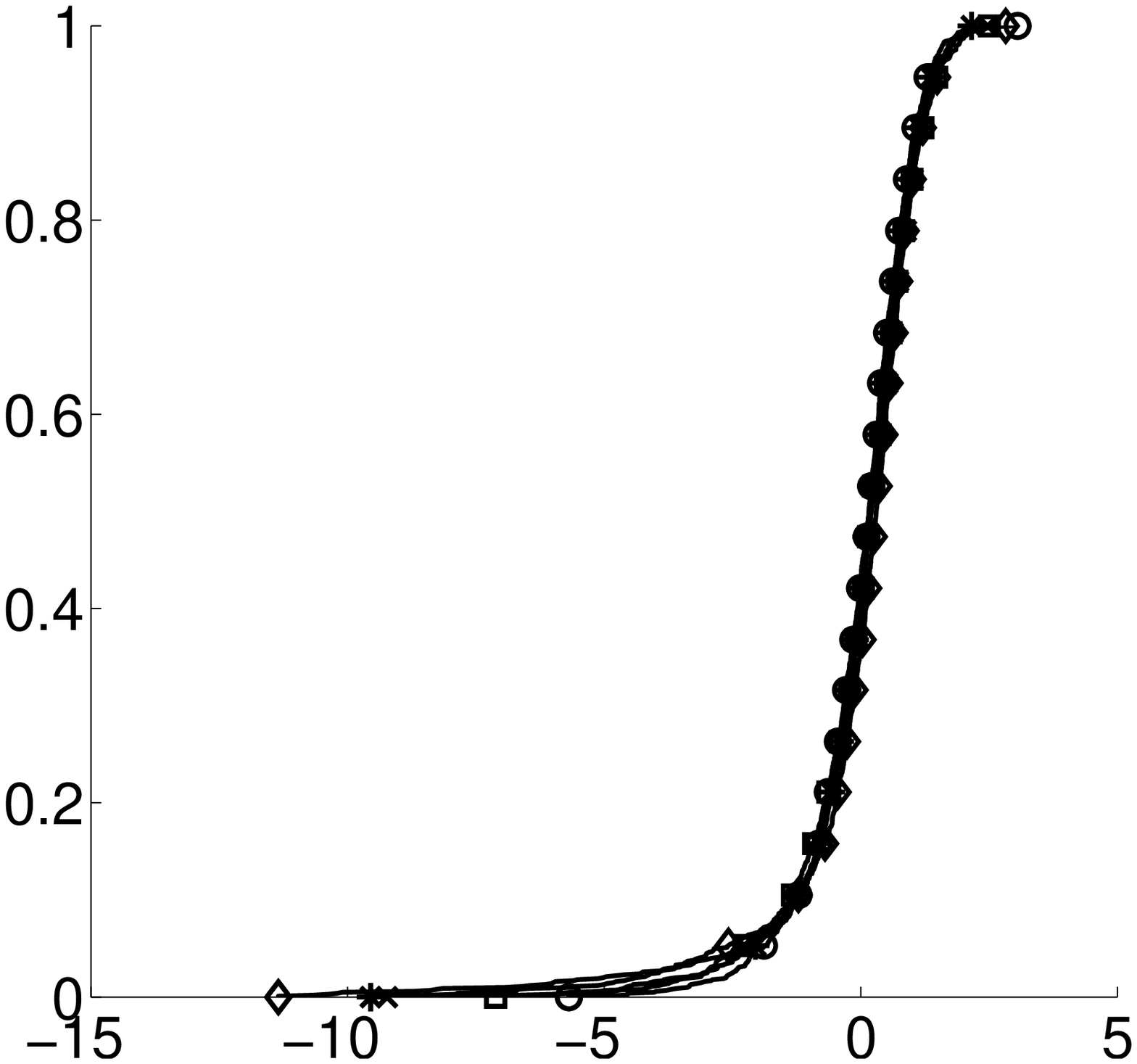}
		\caption{III - $N^{-1/2}$}
	\end{subfigure}
	\caption{Plots of the empirical cumulative distribution function of $\rho_+^-$ using different 
	scaling and for different choices of $\gamma_\pm$. The left column is scaled by $N^{1/\gamma_{
	\text{min}} - 1}$, the center column by $N^{2/\gamma_+ + 2/\gamma_- - 3}$ and the
	right column by $N^{-1/2}$. The first row is for ECM graphs with $\gamma_\pm = 1.3$, the
	second for $\gamma_+ = 1.9$, $\gamma_- = 1.3$ and the third for $\gamma_+ = 1.9$, $\gamma_-
	= 1.5$, corresponding to points I, II and III, respectively, in Figure~\ref{fig:scaling_area}.}
	\label{fig:phase_transitions}
\end{figure}

\begin{table}%
\centering
\begin{tabular}{ccc}
	Region &  & $f(\gamma_+, \gamma_-)$ \\
	\hline
	A & & $1/\gamma_{\text{min}} \, - 1$ \\
	B & & $(2/\gamma_+) + (2/\gamma_-) - 3$ \\
	C & & $-1/2$
\end{tabular}
\caption{The three scaling terms for $\rho_+^-$ for each of the three regions, displayed in Figure
\ref{fig:scaling_area}}
\label{tbl:scaling_terms}
\end{table}

\subsection{Phase transitions for the Out-In degree-degree dependency}

First we remark that for the CM, the empirical distribution of the degrees on both sides of a randomly 
sampled edge converges to the distribution of two independent random variables as $N^{-1}$, see 
\cite{Hoorn2014}. Because Spearman's rho and Kendall's tau on independent joint measurements are normal 
statistics~\cite{Hoeffding1948}, the scaling of their average is $N^{-1/2}$. Hence $\rho_\alpha^\beta$ 
for CM graphs scales as 
$N^{-1/2}$. Since an ECM graph is basically a CM graph where multiple edges are merged and self-loops 
are removed, it follows that the distributions for the degrees on both side of a randomly chosen edge 
differ from 
those of the CM by terms of the order $\sum_{i,j = 1}^N E_{ij}^c/E$. Therefore, the scaling of 
$\rho_+^-$ is determined by the largest term out of $N^{-1/2}$ and the scaling of $\sum_{i,j = 1}^N 
E_{ij}^c/E$. Since the latter undergoes a phase transition, we actually have a three stage phase 
transition for the scaling of $\rho_+^-$ in the ECM. The first stage has scaling $N^{-1 + 
1/\gamma_{\text{min}}}$ and holds for all $\gamma_\pm$ for which 
\[
	\frac{1}{\gamma_{\text{min}}} - 1 \le \frac{2}{\gamma_+} + \frac{2}{\gamma_-} - 3,
\]
since both correspond to upper bounds. The next region, $\gamma_\pm$ such that $2/\gamma_+ + 2/\gamma_- - 
3 \ge -1/2$, has scaling $N^{2/\gamma_+ + 2/\gamma_- - 3}$. Outside this region we have normal scaling, 
$N^{-1/2}$. The different regions are displayed in Figure~\ref{fig:scaling_area}, while Table
\ref{tbl:scaling_terms} shows the three scaling terms. We remark that the phase transitions of the 
scaling are smooth since they are induced by inequalities on the terms.

\begin{figure}[htp]
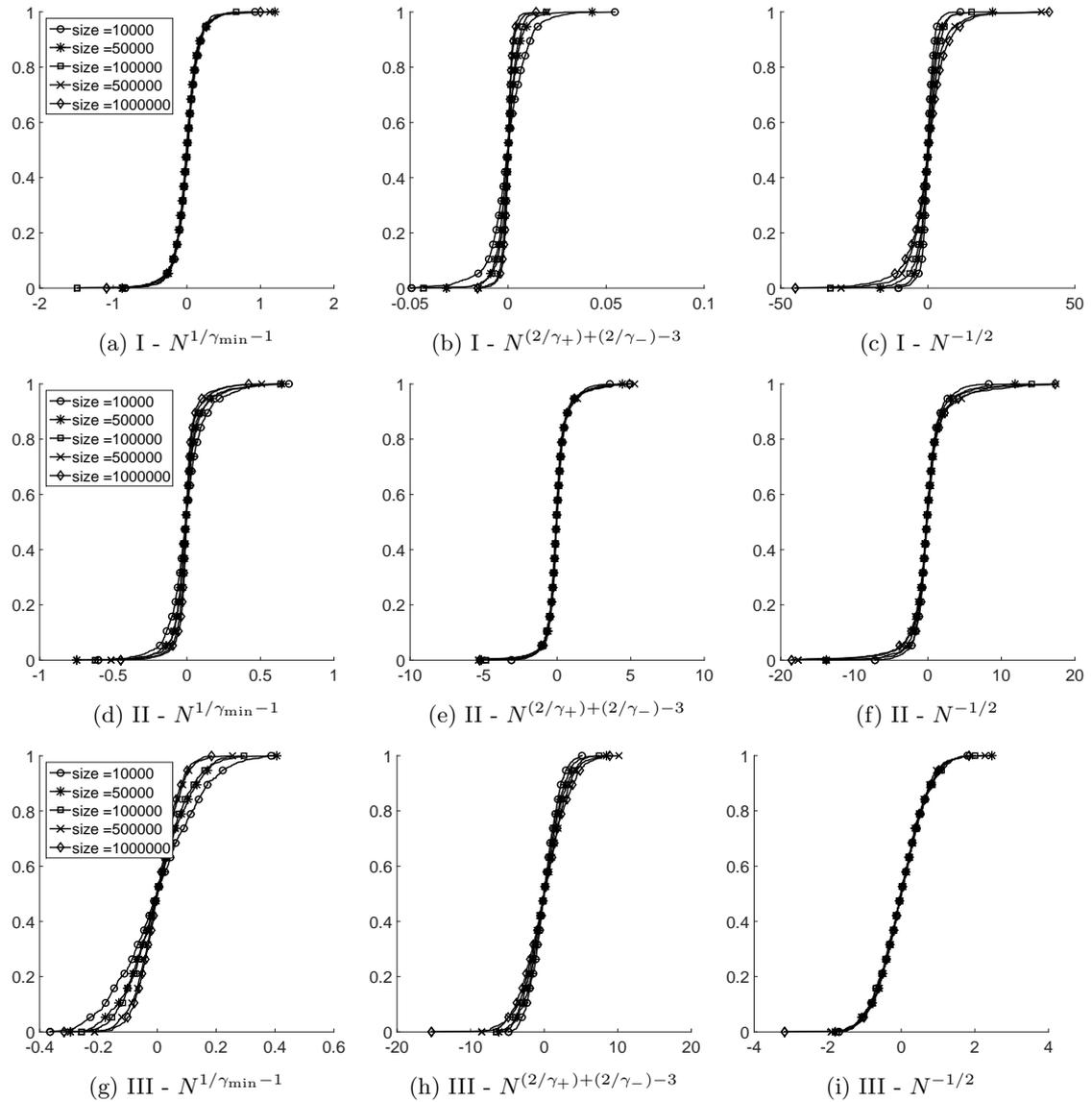
%
	\begin{subfigure}{0.32\linewidth}
		\includegraphics[scale=0.26]{figures_scaling_out_out_1313_slln.eps}
		\caption{I - $N^{1/\gamma_{\text{min}} - 1}$}
		\vspace{5pt}
	\end{subfigure} 
	\begin{subfigure}{0.32\linewidth}
		\includegraphics[scale=0.26]{figures_scaling_out_out_1313_poisson.eps}
		\caption{I - $N^{(2/\gamma_+) + (2/\gamma_-) - 3}$}
		\vspace{5pt}
	\end{subfigure}
	\begin{subfigure}{0.32\linewidth}
		\includegraphics[scale=0.26]{figures_scaling_out_out_1313_sqrt.eps}
		\caption{I - $N^{-1/2}$}
		\vspace{5pt}
	\end{subfigure}
	\begin{subfigure}{0.32\linewidth}
		\includegraphics[scale=0.26]{figures_scaling_out_out_1319_slln.eps}
		\caption{II - $N^{1/\gamma_{\text{min}} - 1}$}
		\vspace{5pt}
	\end{subfigure} 
	\begin{subfigure}{0.32\linewidth}
		\includegraphics[scale=0.26]{figures_scaling_out_out_1319_poisson.eps}
		\caption{II - $N^{(2/\gamma_+) + (2/\gamma_-) - 3}$}
		\vspace{5pt}
	\end{subfigure}
	\begin{subfigure}{0.32\linewidth}
		\includegraphics[scale=0.26]{figures_scaling_out_out_1319_sqrt.eps}
		\caption{II - $N^{-1/2}$}
		\vspace{5pt}
	\end{subfigure}
	\begin{subfigure}{0.32\linewidth}
		\includegraphics[scale=0.26]{figures_scaling_out_out_1519_slln.eps}
		\caption{III - $N^{1/\gamma_{\text{min}} - 1}$}
	\end{subfigure} 
	\begin{subfigure}{0.32\linewidth}
		\includegraphics[scale=0.26]{figures_scaling_out_out_1519_poisson.eps}
		\caption{III - $N^{(2/\gamma_+) + (2/\gamma_-) - 3}$}
	\end{subfigure}
	\begin{subfigure}{0.32\linewidth}
		\includegraphics[scale=0.26]{figures_scaling_out_out_1519_sqrt.eps}
		\caption{III - $N^{-1/2}$}
	\end{subfigure}
	\caption{Plots of the empirical cumulative distribution function of $\rho_+^+$ for choices of 
	$\gamma_\pm$ corresponding to points I, II and III from	Figure~\ref{fig:scaling_area}, 
	using	the corresponding scaling. }
	\label{fig:phase_transitions_outout}
\end{figure}

\subsection{Simulations}

In order to show the phase transitions we plotted the empirical cumulative distribution function of
$\rho_+^-$ for the specific choices of $\gamma_\pm$, corresponding to the points I, II and III in
Figure~\ref{fig:scaling_area}. For each of the three points we shifted the empirical data by its 
average and multiplied it by $N^{-f(\gamma_+, \gamma_-)}$, for any of the three coefficients from
Table~\ref{tbl:scaling_terms}, corresponding to the different scaling areas A, B and C. The results 
are shown in Figure~\ref{fig:phase_transitions}. When the correct scaling is applied, the 
corresponding cdf plots should almost completely overlap and resemble the cdf of some limiting 
distribution. We observe that for each of the three choices I, II and III, this is the case when the 
corresponding scaling from its area, respectively A, B and C, is chosen.

%% file: figures_scaling_scalingarea.tex
\begin{tikzpicture}
	
	\draw (0,0) -- (0,5);
	\draw (0,0) -- (5,0);
	\draw (5,0) -- (5,5);
	\draw (5,5) -- (0,5);
	\draw (0,0) node[below] {$1$};
	\draw (5,0) node[below] {$2$};
	\draw (0,0) node[left] {$1$};
	\draw (0,5) node[left] {$2$};
	
	\draw (2.5,-0.3) node {$\gamma_+$};
	\draw (-0.3,2.5) node {$\gamma_-$};
	
	\draw[black,domain=0:2.5] plot (\x, {(2*\x + 10)/(0.4*\x + 1) - 5});
	\draw[black,domain=2.5:5] plot (\x, {(\x + 5)/(0.4*\x) - 5});
	\draw[black,domain=1.6667:5] plot (\x, {(4*\x + 20)/(\x + 1) - 5}); 
	
	\draw (1,0.5) node {$A$};
	\draw (1,4.4) node {$B$};
	\draw (4.4,4.4) node {$C$};
	
	\draw (1.5,1.5) node[above left] {I};
	\draw (1.5,1.5) node[draw,circle,inner sep=1pt,fill] {};
	\draw (4.5,1.5) node[above left] {II};
	\draw (4.5,1.5) node[draw,circle,inner sep=1pt,fill] {};
	\draw (4.5,2.5) node[above left] {III};
	\draw (4.5,2.5) node[draw,circle,inner sep=1pt,fill] {};

\end{tikzpicture}

%% file: scaling_other_cases.tex
\section{Scaling of degree-degree dependencies for the other cases}
\label{sec:other_cases}

\begin{figure}[htp]%
	\begin{subfigure}{0.32\linewidth}
		\includegraphics[scale=0.32]{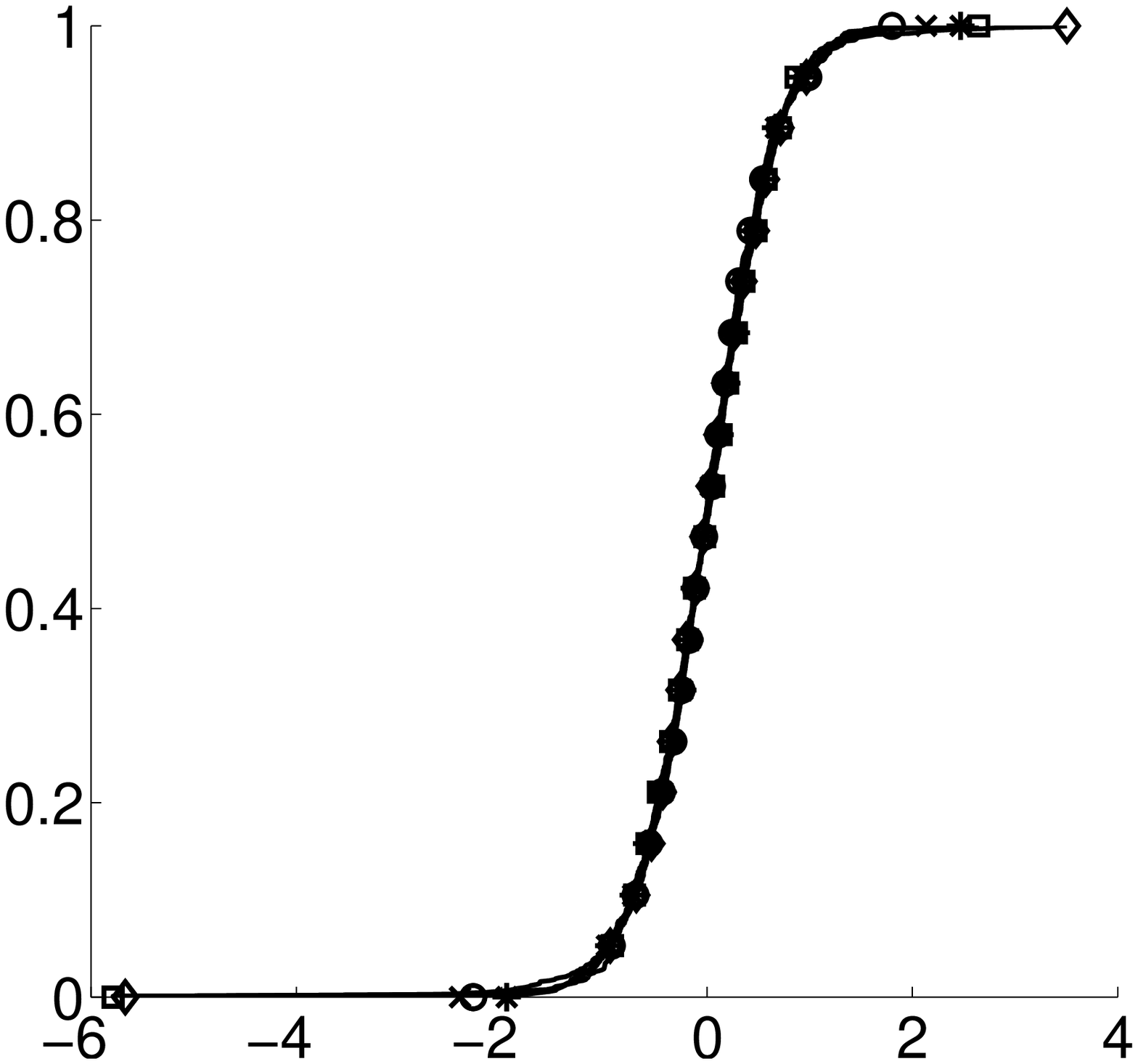}
		\caption{I - $N^{-1/2}$}
	\end{subfigure} 
	\begin{subfigure}{0.32\linewidth}
		\includegraphics[scale=0.32]{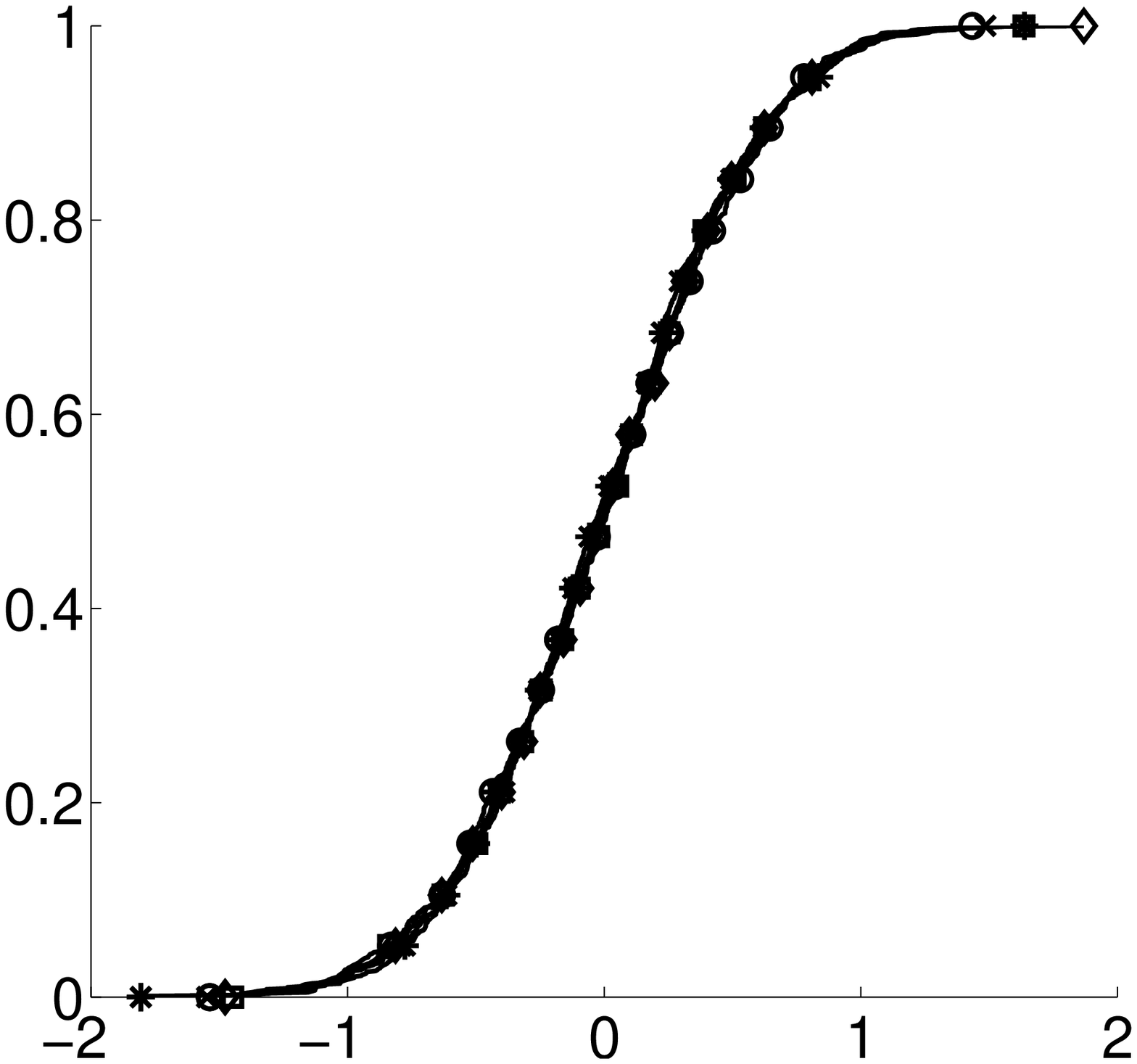}
		\caption{II - $N^{-1/2}$}
	\end{subfigure}
	\begin{subfigure}{0.32\linewidth}
		\includegraphics[scale=0.32]{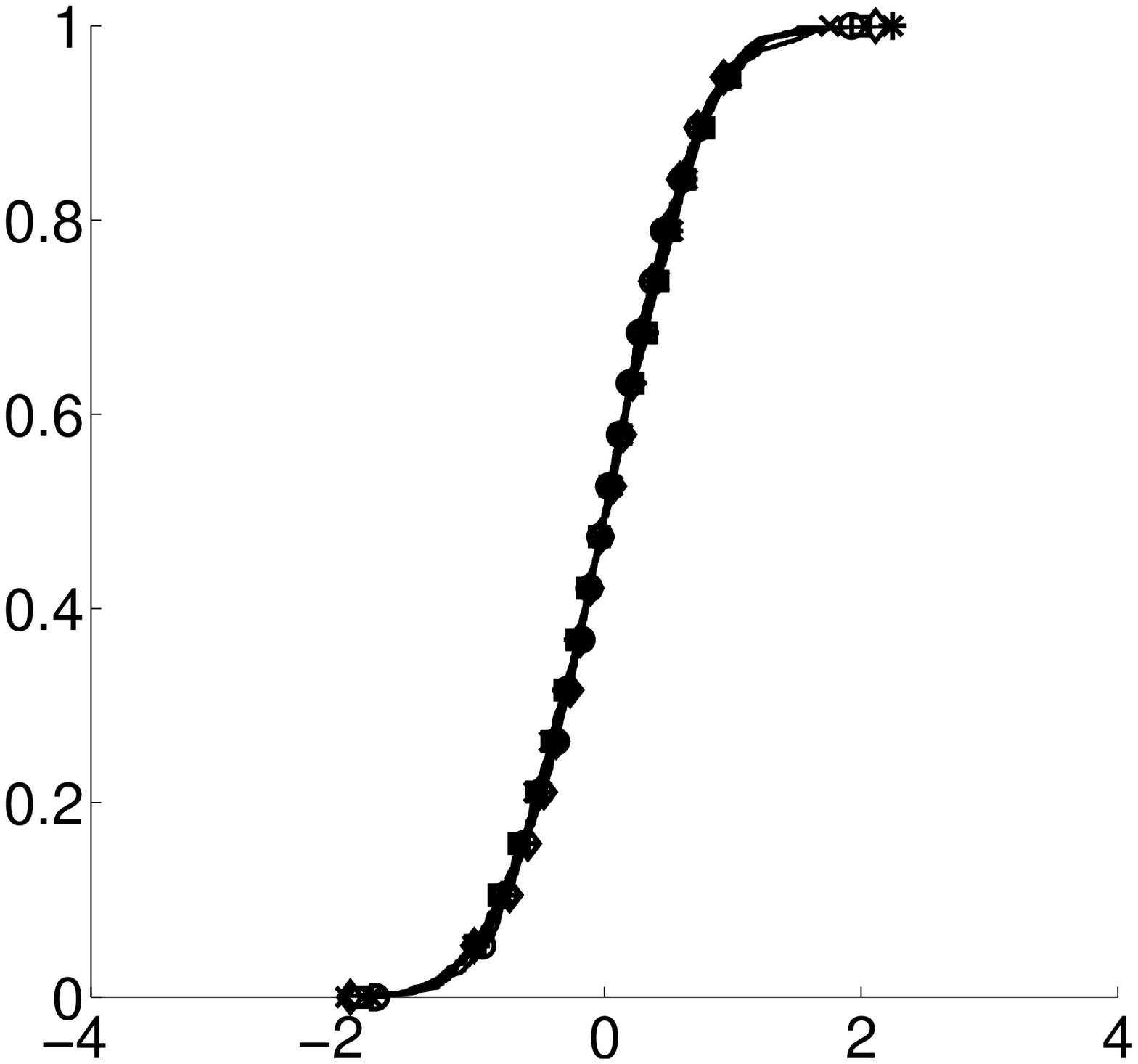}
		\caption{III - $N^{-1/2}$}
	\end{subfigure}
	\caption{Plots of the empirical cumulative distribution function of $\rho_-^+$ for choices of 
	$\gamma_\pm$ corresponding to points I, II and III from	Figure~\ref{fig:scaling_area}, 
	using	square root scaling. }
	\label{fig:scaling_in_out}
\end{figure}

In the previous section we completely characterized the scaling behavior of $\rho_+^-$ for ECM graphs 
with infinite variance of the degrees. Here, we first discuss the remaining correlation types, 
$\rho_+^+$, $\rho_-^-$ and $\rho_-^+$ in the infinite variance regime and lastly, we consider all 
four types in the finite variance regime.

The intuition behind the structural negative Out-In dependencies was that multiple edges are more
likely to exist between nodes of large out- and in-degree. The other three types do not show
negative correlations, see Figure~\ref{sfig:neg_correlations_inout}-\ref{sfig:neg_correlations_inin},
which we argued was due to the fact that the in- and out-degree of a node in the ECM are independent.
Nevertheless, the spread of both the Out-Out and In-In degree-degree dependency exhibits scaling
with the same functions as the Out-In dependency. This is illustrated in Figure
\ref{fig:phase_transitions_outout}, where we plotted the empirical cumulative distribution of the 
Out-Out dependency for ECM graphs, for values of $\gamma_\pm$ corresponding to points I, II and III 
from Figure~\ref{fig:scaling_area}, scaled by the correct term for each of these points.  This is 
because $\rho_+^+$ again depends on the number of erased edges, through the out-degree of their source 
nodes. However, the out-degree of the target node of a removed edge can be both large or small, thus 
$\rho_+^+$ in the ECM remains zero on average. By symmetry, the scaling for the In-In dependency is 
similar. 

This non-trivial scaling is typical for the ECM. Recall that in the CM, $\rho_\alpha^\beta$ is a 
normal statistic and scales as $N^{-1/2}$ for any $\alpha,\beta$ because all degrees are independent 
random variables. This is exactly what we observe for the In-Out degree-degree dependency, which, in 
contrast to the other three, is not biased towards removed edges. As we expect, here we have normal, 
square root, scaling for ECM graphs for any choice of $\gamma_\pm$. This can clearly be observed in 
Figure~\ref{fig:scaling_in_out}, where we plotted the empirical cumulative distributions of 
$\rho_-^+$ scaled by $N^{-1/2}$.

\begin{figure}[t]%
	\begin{subfigure}{.5\linewidth}
		\includegraphics[scale=0.35]{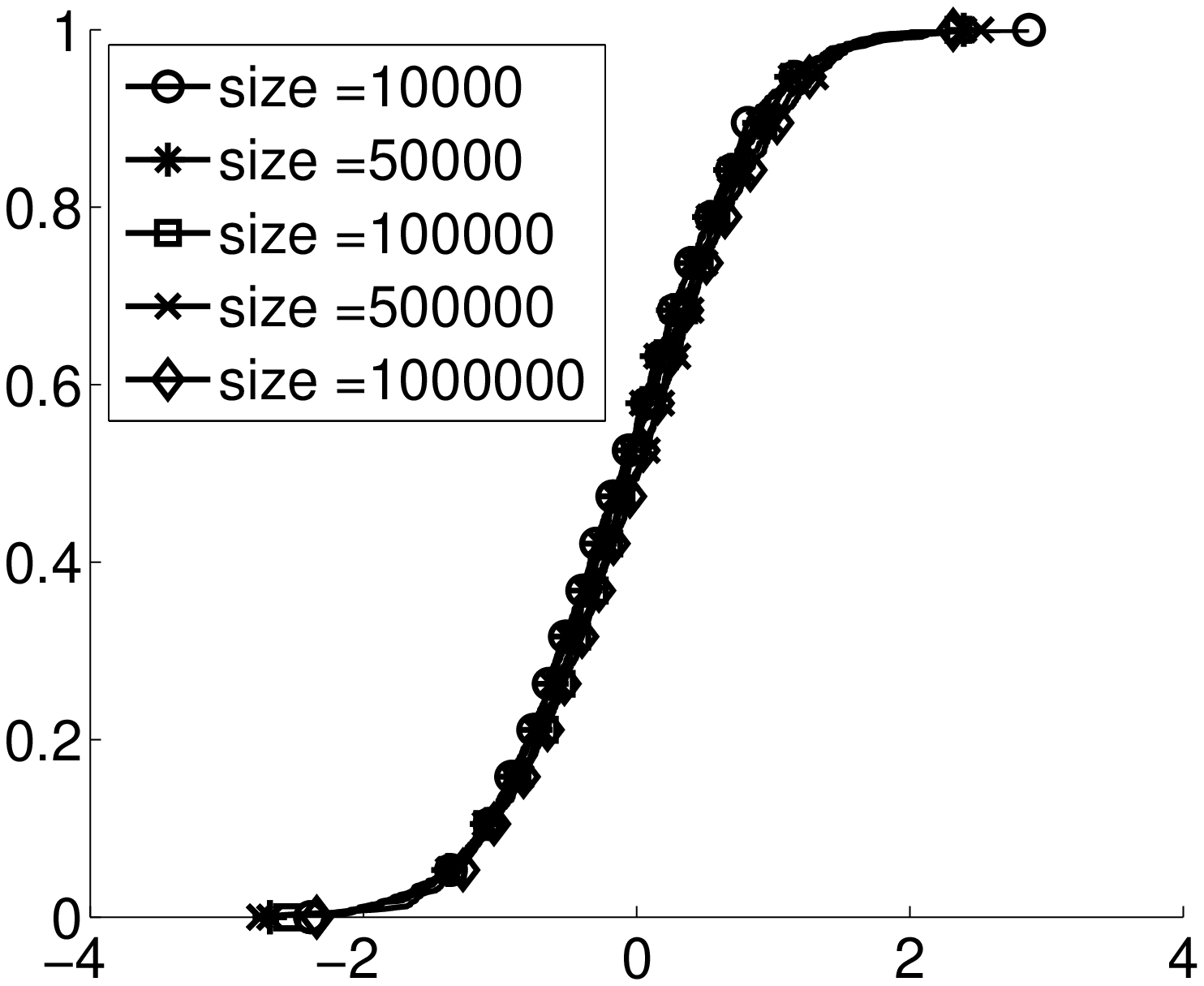}
		\caption{Out-In}
		\label{sfig:normal_scaled_corr_outin}
	\end{subfigure}~
	\begin{subfigure}{.5\linewidth}
		\includegraphics[scale=0.35]{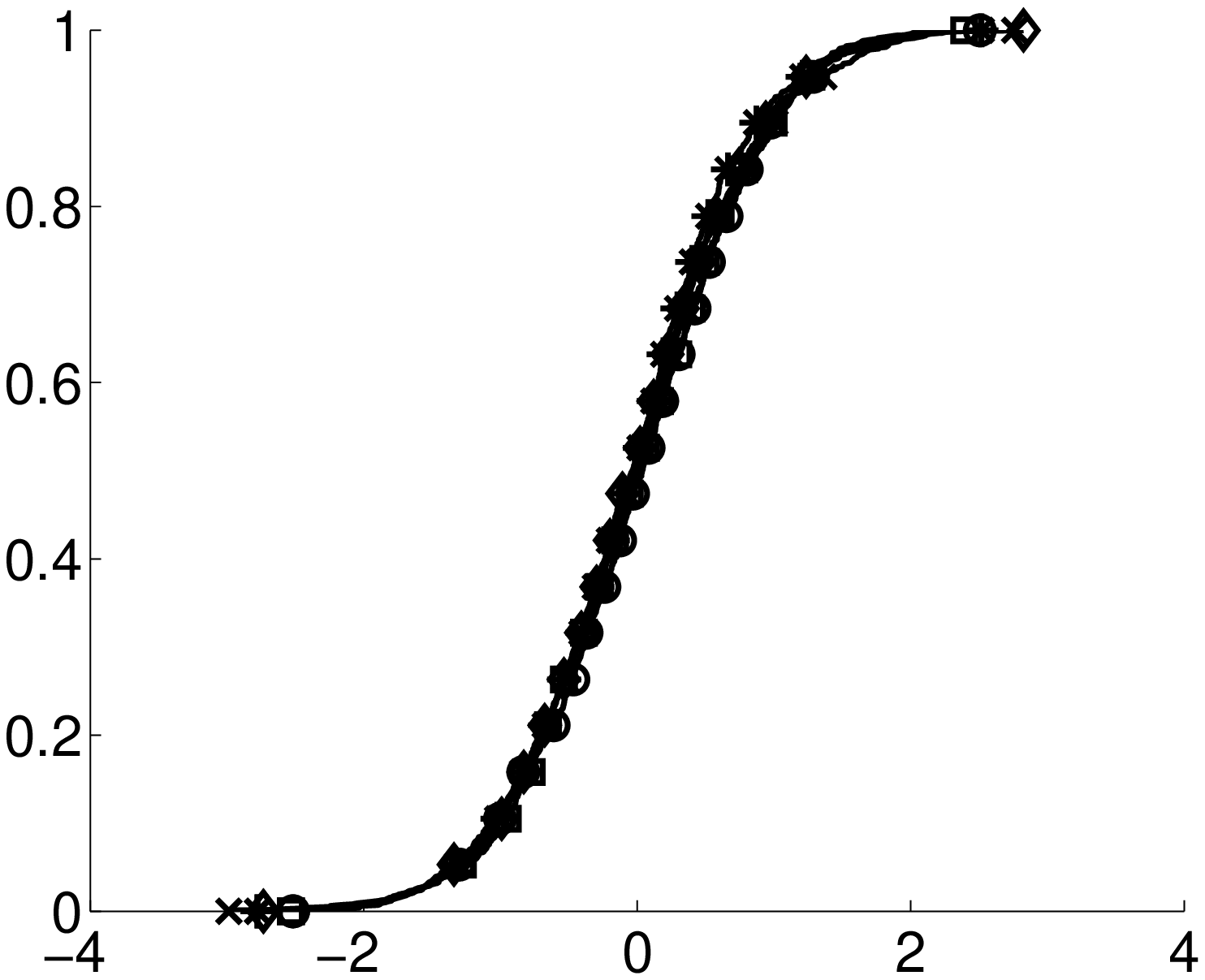}
		\caption{In-Out}
		\label{sfig:normal_scaled_corr_inout}
	\end{subfigure}
	
	\begin{subfigure}{.5\linewidth}
		\includegraphics[scale=0.35]{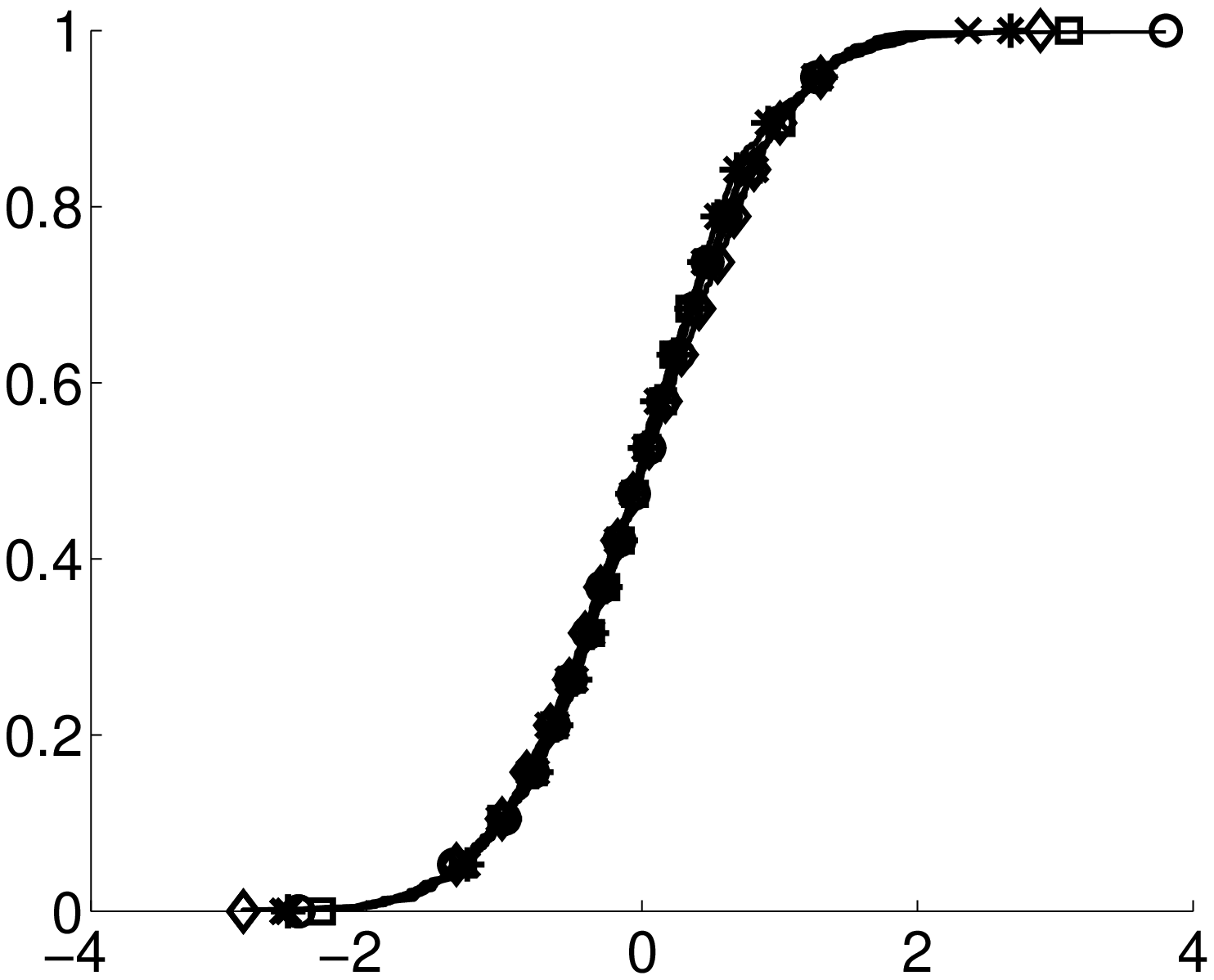}
		\caption{Out-Out}
		\label{sfig:normal_scaled_corr_outout}
	\end{subfigure}~
	\begin{subfigure}{.5\linewidth}
		\includegraphics[scale=0.35]{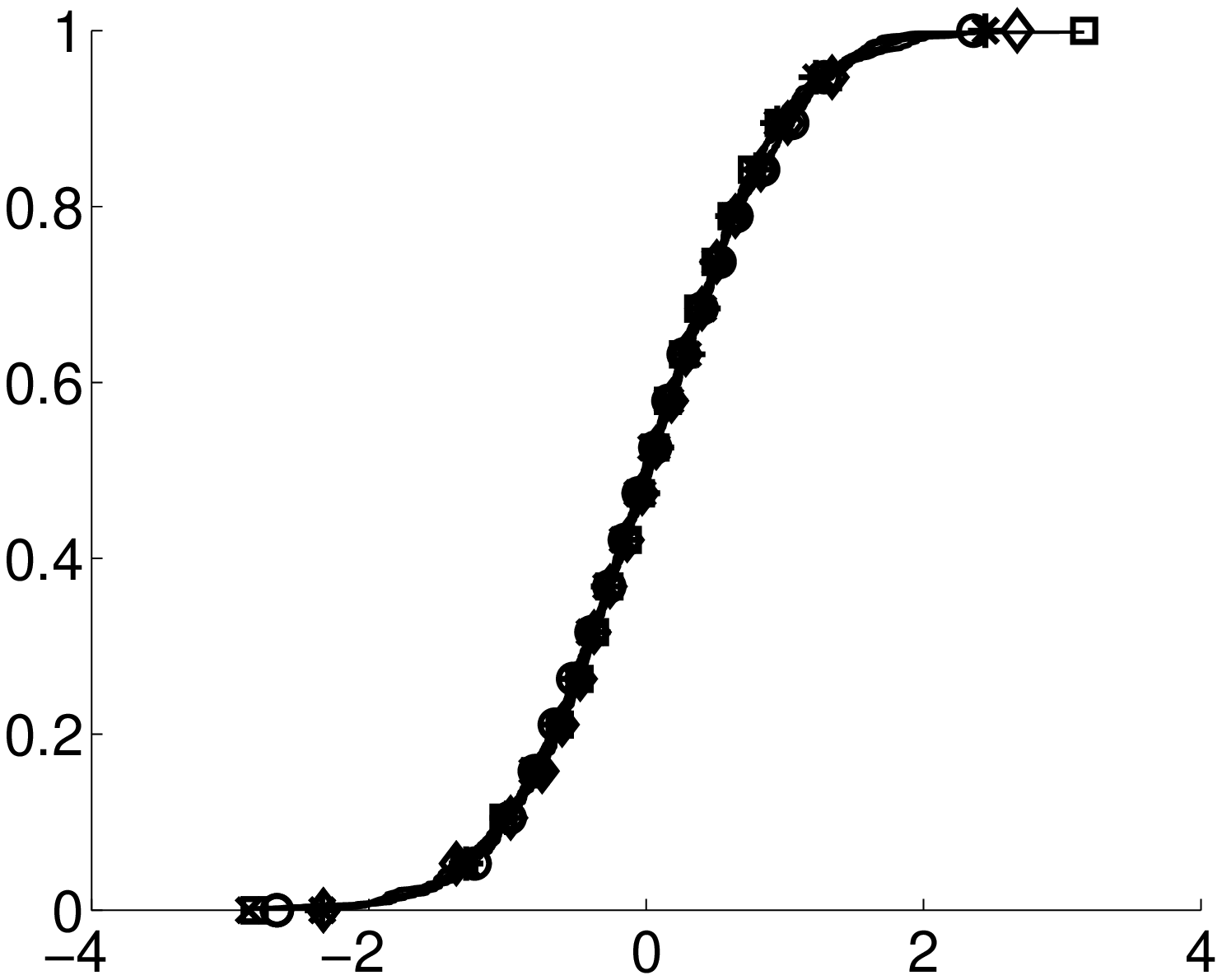}
		\caption{In-In}
		\label{sfig:normal_scaled_corr_inin}
	\end{subfigure}
	\caption{Plots of the empirical cumulative distribution of $\rho_\alpha^\beta$ for all four 
	degree-degree dependency types for ECM graphs with $\gamma_\pm = 2.1$ of different sizes, 
	scaled by $N^{-1/2}$. Each plot is based on $10^3$ realizations of the model.}
	\label{fig:normal_scaled_correlations_example}%
\end{figure}

\begin{figure}%
	\centering
	\includegraphics[scale=.5]{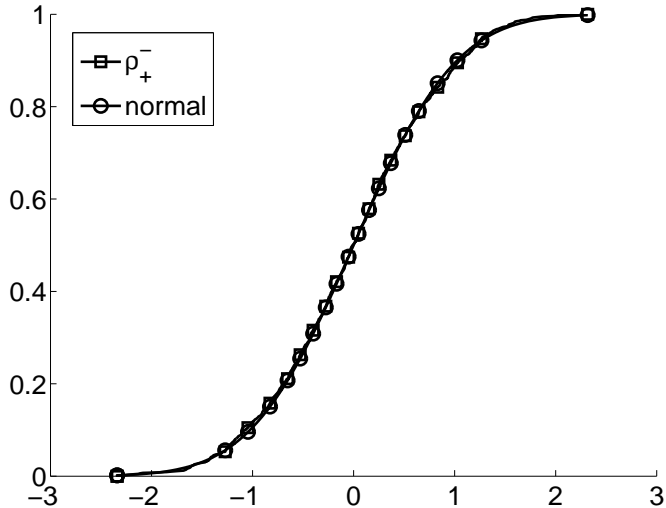}%
	\caption{Plot of the empirical cumulative distribution function of $\rho_+^-$ for ECM graphs of 
	size $10^6$ with $\gamma_\pm = 2.1$ and a normal cumulative distribution with $\mu = 0$	and 
	$\sigma^2 = 0.8$.}%
	\label{fig:2121_normal_scaling}%
\end{figure}

For the degree-degree dependencies in the finite variance regime we plotted the empirical cumulative
distributions of $\rho_\alpha^\beta$, scaled by $N^{-1/2}$, in Figure
\ref{fig:normal_scaled_correlations_example}. Since these are all completely similar, we took the 
plot for $\rho_+^-$ for an ECM graph of size $10^6$ and compared it to a fitted normal distribution 
with $\mu = 0$ and $\sigma^2 = 0.8$, see Figure~\ref{fig:2121_normal_scaling}. These plots strongly 
overlap enforcing the claim that for ECM graphs with finite degree variance all four correlations are 
normal statistics. 

%% file: conclusion.tex
\section{Conclusion and Discussion}

In this paper we analyzed degree-degree dependencies in the directed Erased Configuration Model. We 
showed, Figure~\ref{fig:negative_correlations_example}, that in the infinite variance regime only 
the Out-In dependency exhibits structural negative values, while all correlations behave similar
when both degrees have finite variance, Figure~\ref{fig:normal_correlations_example}. We 
investigated the scaling of the structural negative Out-In correlations. These undergo a phase 
transition in terms of the exponents $\gamma_\pm$ of the degree distributions
(\ref{eq:in_out_distribution}), which we showed by establishing two upper bounds, 
(\ref{eq:first_scaling_bound}) and~(\ref{eq:second_scaling_bound}), on the total average removed 
number of edges, both of which scale at different rates. Combining this with the square root
scaling of Spearman's rho and Kendall's tau, we identified three regions, depending on $\gamma_\pm$,
with different scaling, Figure~\ref{fig:scaling_area}, and illustrated their phase transitions in 
Figure~\ref{fig:phase_transitions}. Next, we considered the remaining three 
dependency types for the infinite variance regime. We showed, Figure~\ref{fig:phase_transitions_outout},
that the scaling of the Out-Out and In-In correlations behaves similarly to the Out-In, even though
they do not exhibit structural negative values, while the In-Out degree-degree dependency has 
square root scaling, Figure~\ref{fig:scaling_in_out}. Finally we investigated the scaling for 
correlations when the degrees have finite variance. In this case all four types have square root
scaling and the plots of the cumulative distributions are very similar, Figure
\ref{fig:normal_scaled_correlations_example}. This was confirmed when we compared the plot of 
$\rho_+^-$ for ECM graphs of size $10^6$, with $\gamma_\pm = 2.1$, with that of a fitted normal 
distribution in Figure~\ref{fig:2121_normal_scaling}.

Our analysis shows that degree-degree dependencies in directed networks display non-trivial 
behavior in terms of scaling when the degrees have infinite variance. This scaling is important
when doing statistical analysis of these measures or their impact on other processes on networks, 
for it determines their spread and hence enables to asses the significance of measurements. 

We showed that degree-degree dependencies for degrees with finite variance, scaled by $N^{-1/2}$, 
converge to a normal distribution with zero mean. We have not yet been able to determine the
variance of these distributions as a function of the tail exponents $\gamma_\pm$ which would
completely characterize their behavior.

For three of the four correlation types in the infinite variance regime, we did not determine 
the limiting distributions. This is mainly due to the fact that we expect these to be  
\emph{stable distributions}, since one of the three scaling regions is due to the Central Limit 
Theorem for regularly varying random variables, hows limits are stable distributions. Although 
these distributions have a well defined characteristic function, their density function, in 
general, does not have an analytical expression. Moreover, we are dealing with discrete data and 
simulation of such distributions is a field of it's own. Nevertheless, we do expect that Central 
Limit Theorems for degree-degree dependencies can be formulated and proven, which would fully 
complete their statistical analysis. 

Finally, our empirical results clearly show the analytically derived phase transitions. However,
the region with the $N^{(2/\gamma_+) + (2/\gamma_-) - 3}$ scaling is less distinct than the other 
two. One of the possible reasons for this is that within the area where this scaling applies, the 
difference in value with the other two terms is small. We therefore picked point II in Figure
\ref{fig:scaling_area} such that this difference was large enough to distinctly show this scaling
visually in the plots.

We close by strongly suggesting to use the ECM as a null model for analysis of degree-degree 
dependencies, both for determining their impact on processes as well as significance. Although for 
the latter, values are often compared to averages, using the rewiring model~\cite{Maslov2002}, we 
emphasize that fixing the degrees imposes strong constraints on the possible simple graphs that can 
be generated. Moreover, in real-life networks, not only wiring but also the degrees of the nodes, 
are a result of a random process. Therefore, in a null-model, it seems more natural to fix only 
general properties of the network, such as degree distributions.